\documentclass[final,3pt,scale=0.8,times,12pt]{elsarticle}
\usepackage{CJK} 
\biboptions{numbers,sort&compress}
\usepackage{graphicx,enumitem} 
\usepackage[margin=0.75in]{geometry}

\usepackage{subfigure}
\usepackage{epstopdf}
\usepackage{bibentry}
\usepackage{caption}
\usepackage{color}
\usepackage{mathrsfs}
\usepackage{multirow}
\usepackage{makecell}
\usepackage{hhline}
\usepackage{lineno,hyperref}
\usepackage{fancyhdr}
\usepackage{geometry}
\usepackage{titlesec}
\usepackage{mathtools}
\usepackage{yhmath}
\usepackage{amsmath}
\hypersetup{
    colorlinks=true,    
    linkcolor=blue,     
    urlcolor=blue,      
    citecolor=green,    
    breaklinks=true     
}
\titleformat{\section}
  {\normalfont\Large\bfseries}{\thesection}{1em}{}
\titleformat{\section}
  {\normalfont\Large\bfseries}{\thesection}{1em}{}
\titleformat{\subsection}
  {\normalfont\bfseries}{\thesubsection}{1em}{} 
\titleformat{\subsubsection}
  {\normalfont\bfseries}{\thesubsubsection}{1em}{}
\DeclareMathOperator{\sech}{sech}
\numberwithin{equation}{section}

\newtheorem{rk}{Remark}
\allowdisplaybreaks[4]

\def\e{{\rm e}}

\def\e{{\rm e}}

\journal{***}
\begin{document}
\begin{frontmatter}

\title{\textbf{Localized stem structures in quasi-resonant solutions of the Kadomtsev-Petviashvili equation}}

\author[a]{Feng Yuan \corref{cor2}}
\author[b]{Jingsong He\corref{cor1}}
\author[c]{Yi Cheng}
\address[a]{College of Science, Nanjing University of Posts and Telecommunications, Nanjing, 210023, P. R. China}
\address[b]{Institute for Advanced Study, Shenzhen University, Shenzhen, 518060, P. R. China}
\address[c]{School of Mathematical Sciences, USTC, Hefei, Anhui 230026, P. R. China}

\cortext[cor1]{Corresponding author. hejingsong@szu.edu.cn}
\cortext[cor2]{Corresponding author. fengyuan@njupt.edu.cn}

\begin{abstract}
When the phase shift of X-shaped solutions before and after interaction is finite but approaches infinity, the vertices of the two V-shaped structures become separated due to the phase shift and are connected by a localized structure, which is referred to as the stem structure. 
This special type of elastic collision is known as a quasi-resonant collision. 
This study investigates quasi-resonant solutions and the associated localized stem structures in the context of the KPII and KPI equations. 
For the KPII equation, we classify quasi-resonant 2-solitons into weakly and strongly types, depending on whether the phase approaches \(-\infty\) or \(+\infty\). 
We analyze their asymptotic forms to detail the trajectories, amplitudes, velocities, and lengths of their stem structures. 
These results of  quasi-resonant 2-solitons are used to to provide  analytical descriptions  of  interesting  patterns of the water waves observed on shallow water surface. 
Similarly, for the KPI equation, we construct quasi-resonant breather-soliton solutions and classify them into weakly and strongly types, based on the behavior of their internal parameters. 
We compare the similarities and differences between the stem structures in the quasi-resonant soliton and the quasi-resonant breather-soliton. 
Additionally, we provide a comprehensive and rigorous analysis of their asymptotic forms and stem structures. 
Our results indicate that the resonant solution, i.e. resonant breather-soliton of the KPI and soliton for the KPII, represents the limiting case of the quasi-resonant solution as phase approaches \(\infty\).
\end{abstract}


\end{frontmatter}
\section{Introduction}
Solitary waves, together with their interaction behaviors, are widespread natural phenomena distributed across oceanic environments, encompassing both the sea surface and multiple subsurface depth layers.
Specifically, oblique interactions between line solitons--referring to intersecting collisions with distinct propagation angles, as opposed to parallel collisions--have been documented in oceanic and laboratory environments.
The investigation of nonlinear waves, crucial to both physics and engineering, has seen significant advancements in both theoretical and experimental domains. 
Nonlinear partial differential equations (NPDEs) that describe the complex evolution of these waves have been extensively studied using various techniques, including the inverse scattering transform \cite{prl1967}, Darboux transformation \cite{dt01,dt02,dt03,dt04}, and Hirota bilinear method \cite{book2004,jpcs2014,nn2022,rao01,rao02}. 
These methodologies have led to the discovery of a variety of nonlinear wave solutions, such as breathers, lumps, and rogue waves.

The Kadomtsev-Petviashvili (KP) equations \cite{kp1970}
\begin{equation}\label{kpeq}
(u_t+6uu_x+u_{xxx})_x+\delta u_{yy}=0,\qquad \delta=\pm 3,
\end{equation}
serve as fundamental models for weakly two-dimensional nonlinear dispersive waves.  
Since their introduction, the KP equations have been derived in several physical contexts, including surface and internal water waves \cite{kp1979}, nonlinear optics \cite{kp1995}, and ion-acoustic waves in plasma under long lateral disturbances \cite{kp2022}.  
Depending on the sign of $\delta$, the system is classified as the KPI equation $(\delta=-3)$ or the KPII equation $(\delta=3)$.  
Although both equations share integrability, they exhibit markedly different solution structures: 
KPI supports breathers and lump-type localized solutions \cite{kpi2016,kpi2017,kpi2021-1,kpspd1983,kpjpsj1989,kp1990,kp1992-1,kp1992-2,kpjmp1998}, 
while KPII admits only stable line solitons, making it an ideal setting for analyzing pure multi-soliton interactions without lump dynamics. 
Exact $N$-soliton solutions of \eqref{kpeq} can be constructed through Hirota bilinear form \cite{hirota1976,kp1976}:
\begin{align}
u^{[n]} &= 2(\ln f^{[n]})_{xx}, \label{equ}\\
f^{[n]} &= \sum\limits_{\mu_i,\mu_j = 0,1} \exp\left( \sum\limits_{i<j}^{N} \mu_i\mu_j A_{ij} + \sum\limits_{i=1}^{N} \mu_i \xi_i \right), \label{nf}
\end{align}
where
\begin{flalign}\label{etakpw}
\begin{split}
\xi_j = k_j x + p_j y + \omega_j t + \xi_j^0,\qquad
\omega_j = -\frac{k_j^4 - \delta p_j^2}{k_j},\\
\exp(A_{ij}) = 
\frac{3k_i^2 k_j^2 (k_i - k_j)^2 + \delta (k_j p_i - k_i p_j)^2}
{3k_i^2 k_j^2 (k_i + k_j)^2 + \delta (k_j p_i + k_i p_j)^2}
\triangleq a_{ij},
\end{split}
\end{flalign}
and $f^{[n]}$ is the Hirota $\tau$-function.  
Moreover, the KP equations possess infinitely many symmetries and conservation laws \cite{kp1982}, and their Cauchy problems have been extensively analyzed under either decaying or line-confined initial data \cite{kp1981,kp1983,kpcmp1990,kp2004,kp2009-1}.

For the KPII equation, line solitons constitute its principal coherent structures, and their interactions display characteristic behaviors such as oblique collisions, resonant Y- and X-type interactions, and the formation of web-like patterns \cite{biondini2007,biondini2006,chakravarty2008}.
Classical studies offered early insight into phase shifts and resonance  during collisions \cite{kp1977,kpjpsj1983}.
Subsequent developments revealed rich algebraic and combinatorial structures underlying KPII line solitons\cite{kp2003,aip2010,kpepj2010,aa2010}, 
including their links to Tamari lattices \cite{jpa2011,Dimakis2011} and their graphic description via totally non-negative Grassmannians \cite{kodama2013}.
Further structural interpretations have been obtained through Young diagrams \cite{jpa2004} and cluster algebra frameworks \cite{kodama2011}.
Numerical investigations have additionally demonstrated the persistence of resonant interactions and the possibility of generating large-amplitude waves through multi-soliton collisions \cite{kp2009-2}. 

The interactions among various solutions of the two classes of KP equations have been extensively studied \cite{kp1978,kpjfm1978,kp2006,mcs2007,kp2007,kp2019,kp2020}, including both elastic and resonant collisions.
Elastic collisions represent a fundamental property of solitons: as solitons pass through one another, they experience a finite phase shift yet recover their original velocities and shapes after interaction \cite{prl1965,cpam1974,book1991}.
The oblique 2-soliton, consisting of two solitons propagating in different (non-parallel) directions that form a specific angle upon collision, undergoes elastic interaction and exhibits an X-type profile. 
This case is called X-shaped soliton. 
In contrast, a qualitatively different phenomenon, referred to as resonance, arises when the wave numbers and frequencies of the interacting solitons satisfy certain conditions.
In resonant collisions, solitons do not return to their initial shapes or velocities; instead, their profiles, amplitudes, and propagation speeds are modified after interaction.
Resonance is characterized by the occurrence of an infinite phase shift \cite{kpjpsj1983}, and the corresponding oblique 2-soliton displays a Y-type structure. 
It is commonly referred to as Y-shaped soliton. 
The KPII equation admits resonant soliton solutions, which have been the subject of intensive study \cite{kpjfm1977,jpa2008,jpa2009,kpii2013,kpii2014-1,kpii2014-2,kpii2015,book2017}. 
Resonant collisions do not occur for the KPI equation when only line solitons are considered.
However, if breathers or lumps are included, resonant interactions become possible \cite{kpi1993,kpi2011}.
The resonant collisions among solitons, breathers, and lumps in the KPI equation have been examined by many authors \cite{kpi2018,kpi2021-2,kpi2022}.
Conversely, KPII does not possess breather or lump solutions, and thus no corresponding resonant structures arise from these types of waves.

Recently, quasi-resonant collisions, which are essentially elastic collisions occurring under conditions where the phase shift is finite but approaches infinity, have obtained attention \cite{jpsj1980,kpjpsj1983,kp2009-2,yuan2024}. 
Notably, quasi-resonant collisions between two oblique solitons can produce localized stem structures. 
During the quasi-resonant process, the single vertex of the X-shaped soliton (which can be regarded as the coincident vertex of two V-shaped solitons) separates due to phase shift, 
giving rise to two discrete V-shaped solitons connected by a local wave and forming the so-called stem structure.
Previous investigations into the stem structures of solitons, although informative, have remained relatively limited in scope and depth.  
For example, Ref.\ \cite{jpsj1980} examined quasi-resonant solitons arising in an extended Boussinesq-like equation, thereby elucidating certain isolated features of stem-type formations.  
More recently, Ref.\ \cite{yuan2024} provided a systematic analysis of stem structures associated with quasi-resonant two-soliton solutions of the asymmetric Nizhnik--Novikov--Veselov (ANNV) system, offering valuable insight into their emergence and qualitative behavior.  
For the KPII equation, quasi-resonant soliton solutions were recognized early within the classical shallow-water framework \cite{kpjpsj1983,kp2009-2}.  
In 2012, Ablowitz and Baldwin \cite{kppre2012} reported field observations of quasi-resonant two-soliton water waves near low tide at two widely separated beaches, interpreted through the KPII model.  
However, their work contained only limited graphical evidence (see Figs.~3--5 in Ref.\ \cite{kppre2012}) and did not provide a detailed examination of the local properties of the intermediate wave, which constitutes a genuine physical realization of a stem structure.  

Despite these advances, several aspects of KPII multi-soliton interactions remain insufficiently understood, in particular the stem, i.e., the local structure connecting two vertices in a multi-soliton interaction pattern.
Kodama's approach can, in principle, determine the locations of such stems, but Ref. \cite{jpa2004} focused on the classification of soliton patterns and did not provide a dedicated analytical description of their structure, although some stem locations were illustrated (e.g., Fig. 2 therein).
Similarly, Ref. \cite{Dimakis2011} employed the tropical approximation to determine the trajectories of all line-soliton branches and noted that, in special cases, additional line-soliton segments induced by phase shifts may appear near branch intersections; these segments correspond precisely to the stem structures considered here (see, for instance, Example 4.3 and Fig. 22 in \cite{Dimakis2011}).
However, no explicit analytical characterization of the graphic and dynamics of the stem in quasi-resonant KPII solutions has yet been carried out in a systematic framework.
The present work therefore aims to provide a detailed analytical and graphical study of the stem structure arising in quasi-resonant soliton (and breather-soliton) interactions of the KP equation, with particular emphasis on its phase graphic and asymptotic configuration.

Conversely, for the KPI equation, quasi-resonant solitons cannot be directly obtained due to the absence of resonant solitons in this model. 
However, previous research has demonstrated that resonant collisions between a breather and a soliton can produce semi-infinite line solitons \cite{bs2021}. 
This leads us to conjecture that quasi-resonant collisions in the KPI equation may also generate stem structures analogous to those observed in the quasi-resonant 2-soliton solutions of the KPII equation \cite{kpjpsj1983,kppre2012}. 
In light of this, the principal focus of this paper is to investigate the stem structures within the quasi-resonant solutions of the KPI and KPII equations using a similar approach as in Ref.\ \cite{yuan2024}. 
Here are the details:
\begin{itemize}
\item Based on the 2-soliton solution of the KPII equation obtained using the Hirota bilinear method, we classify quasi-resonant collisions into weakly quasi-resonant collisions and strongly quasi-resonant collisions, depending on whether \(a_{12} \approx 0\) or \(a_{12} \approx +\infty\). 
Using the asymptotic forms for these scenarios, we conduct a thorough investigation about the properties of their stem structures, including trajectories, amplitudes, velocities, and lengths. Our findings reveal that resonant solitons represent the limiting case of quasi-resonant solitons as \(\epsilon \to 0\). 
These analytical results are used to describe the patterns observed on shallow water surface, such as Venice Beach \cite{kppre2012,photo,jpa2014}.
\item Similarly, based on the 3-soliton solution of the KPI equation given by the Hirota bilinear method, we construct quasi-resonant breather-soliton solutions. 
We further classify these quasi-resonant breather-solitons into weakly and strongly cases based on whether \(\alpha_1^2 + \beta_1^2 \approx 0\) or \(\alpha_1^2 + \beta_1^2 \approx +\infty\) (see \eqref{alphabeta}, equivalent to \(a_{13} \approx 0\) or \(a_{13} \approx +\infty\)). 
The asymptotic forms of these two quasi-resonant scenarios are analyzed, and the properties of their stem structures are examined in detail. 
Additionally, by varying the parameter \(\epsilon\), we demonstrate that resonant breather-solitons represent the limiting case of quasi-resonant breather-solitons as \(\epsilon \to 0\).
\end{itemize}

The paper is organized as follows: Section \ref{sec2} explores the specific properties of the stem structure of the quasi-resonant soliton of the KPII equation, categorizing them into weakly and strongly cases, and uses these categories to describe the two V-shaped waves connected by one stem structure off the coast. 
Section \ref{sec3} constructs the quasi-resonant breather-soliton solution and investigates the associated the stem structure, also the quasi-resonant solution is divided into weakly and strongly cases. 
Finally, Section \ref{summary} offers a summary and discussion of our findings.

\section{Stem structure in the quasi-resonant 2-soliton of KPII equation}\label{sec2}
In this section, we focus on the stem structure in the quasi-resonant 2-soliton of the KPII equation \eqref{kpeq} with $\delta=3$. 
By setting $\delta=3$ and $n=2$ in \eqref{nf}, the tau function of the 2-soliton is expressed as
\begin{equation}
f^{[2]}=1+\exp\xi_1+\exp\xi_2+a_{12}\exp(\xi_1+\xi_2).\label{2f}
\end{equation}
Then 2-soliton solution of the KPII equation is then given by \eqref{2f} and \eqref{equ}, and the smoothness condition is given by $a_{12}\geqslant 0$ ($a_{12}=0$ means the limit $a_{12}\to 0$). 
When the phase shift (denoted as $\Delta_{12}=\ln a_{12}$) is finite, two soliton undergoes an elastic collision and takes on an X-shape. 
While the phase shift is infinite, the 2-soliton undergoes a resonant collision and becomes Y-shape which is given by Eqs.\ \eqref{weakresonance} and \eqref{strongresonance}. 
Interestingly, there is an intermediate state between these two types of collisions: 
when $\Delta_{12}$ is finite but approaching infinity which can be implemented by setting very small value or very large value of $a_{12}$, or denote shortly it by $a_{12}\approx 0$ or $a_{12} \approx \infty$, the 2-soliton undergoes a quasi-resonant collision. 
At this state, the vertices of the X-shaped soliton separate, creating a local stem structure connecting the vertices of the two V-shaped solitons \cite{kpjpsj1983,kppre2012,yuan2024}. 
We refer to the case where $a_{12}\approx 0$ as a weakly quasi-resonant collision, and the case where $a_{12}\approx +\infty$ as a strong quasi-resonant collision. Below, we discuss the properties of the stem structure in both cases.

\subsection{Stem structure in weakly quasi-resonant soliton}\label{sec2.1}
In the scenario where $a_{12}\approx 0$ ($\Delta_{12}\approx -\infty$), the 2-soliton undergoes weakly quasi-resonant collisions. 
To ensure \( a_{12} \approx 0 \), we must choose \( p_2 = \frac{k_2(k_1^2 - k_1k_2 + p_1)}{k_1} - \epsilon \) or \( p_2 = -\frac{k_2(k_1^2 - k_1k_2 - p_1)}{k_1} + \epsilon \), where \( \epsilon \approx 0 \) is a sufficiently small number. 
Substituting these expressions into \( a_{12} \), we obtain $a_{12} = 1 - \frac{4k_1k_2^3}{4k_1k_2^3 + 2\epsilon k_2(k_1 - k_2) - \epsilon^2}$. 
It is evident that if \( k_1 = k_2 \), then \( a_{12} < 0 \). To ensure the smoothness of the solution, we consider the following five cases: 
(1) \( k_1 > k_2 > 0, \epsilon > 0 \); (2) \( k_2 > k_1 > 0, \epsilon < 0 \); (3) \( k_1k_2 < 0, \epsilon > 0 \); (4) \( k_1 < k_2 < 0, \epsilon > 0 \); (5) \( k_2 < k_1 < 0, \epsilon < 0 \). 
Without loss of generality, we will focus on the case (1) where \( p_2 = \frac{k_2(k_1^2 - k_1k_2 + p_1)}{k_1} - \epsilon \), \( k_1 > k_2 > 0 \), and \( \epsilon > 0 \) in this section \ref{sec2.1}.
\begin{rk}
In this paper, $\epsilon$ represents a real constant approximately equal to zero. 
For convenience, we take $\epsilon<10^{-2}$ in the figures to make the stem structure more visible.
\end{rk}
\begin{rk}
Each parameter in the formulas throughout the paper must satisfy the corresponding conditions (quasi-resonant or resonant contidition).
\end{rk}

\begin{figure}[h!tb]
	\centering
    \subfigure[3D plot]{\includegraphics[height=4cm,width=4cm]{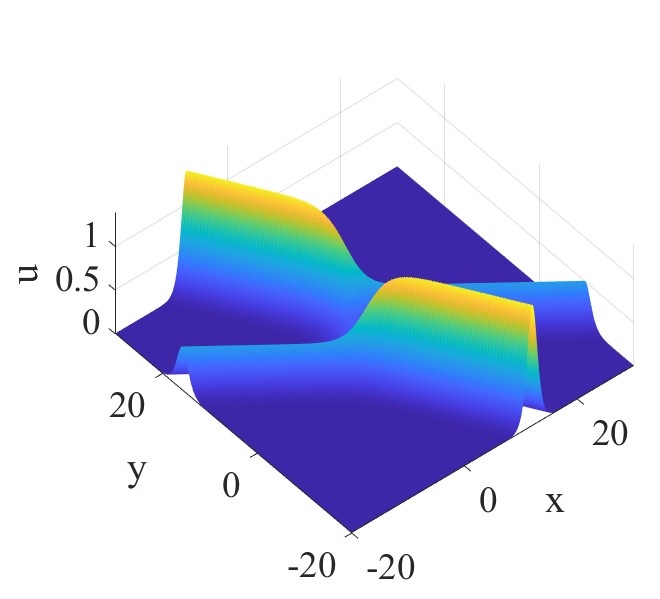}}
    \subfigure[Density map]{\includegraphics[height=4cm,width=5cm]{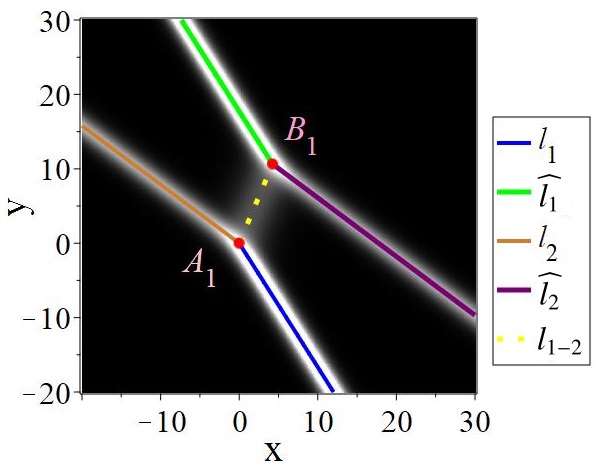}}
    \subfigure[Section curve $u|_{l_{1-2}}$]{\includegraphics[height=4cm,width=4cm]{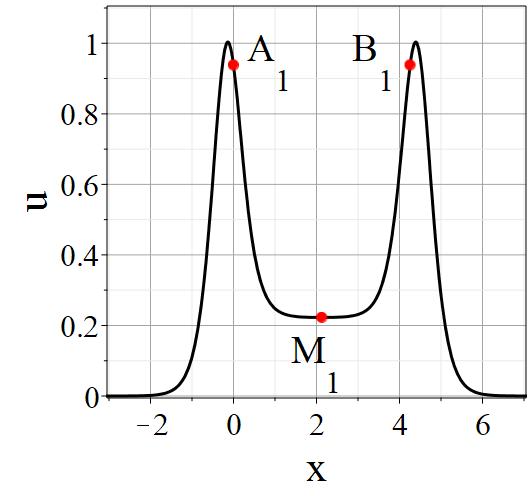}}
	\caption{The weakly quasi-resonant 2-soliton $u_{qw}^{[2]}$ with $k_1=\frac{5}{3},\,k_2=1,\,p_1=1,\,p_2=\frac { k_2(k_1^{2}-k_1k_2+ p_1)}{k_1}-\epsilon,\,\epsilon=10^{-7}
,\,t=0$. (a) 3D map; (b) The density plot and trajectories; (c) The section-cross curve $u|_{l_{1-2}}$.}\label{fig2s-1}
\end{figure}

In order to distinguish two quasi-resonant cases, we denote the tau function of the weakly quasi-resonance as
\begin{equation}\label{2f1}
f_{qw}^{[2]}=1+\e^{\xi_1}+\e^{\xi_2}+a_{12}\e^{\xi_1+\xi_2}.
\end{equation}
And then the weakly quasi-resonant 2-soliton is given by $u_{qw}^{[2]}=2(\ln f_{qw}^{[2]})_{xx}$. 
Based on the asymptotic analysis method given in Refs.\ \cite{jpsj1980,kpjpsj1983}, the weakly quasi-resonant 2-soliton, which is depicted in Fig. \ref{fig2s-1} (a), has four arms and a stem structure and their asymptotic forms are as follwing:\\
Before collision:
\begin{flalign}
\begin{split}
&S_1\,(\xi_1\approx 0,\,\xi_2\to -\infty):\,f\sim 1+\e^{\xi_1},\,u\sim u_{1}=\frac{k_1^2}{2}\sech^2\bigg(\frac{\xi_1}{2}\bigg),\\
&S_2\,(\xi_2\approx 0,\,\xi_1\to -\infty):\,f\sim 1+\e^{\xi_2},\,u\sim u_{2}=\frac{k_2^2}{2}\sech^2\bigg(\frac{\xi_2}{2}\bigg);\\
\end{split}\label{xasy01}
\end{flalign}
After collision:
\begin{flalign}
\begin{split}
&S_1\, (\xi_1+\ln a_{12}\approx 0,\,\xi_2\to +\infty):\,f\sim 1+a_{12}\e^{\xi_1},\,u\sim \widehat{u_{1}}=\frac{k_1^2}{2}\sech^2\bigg(\frac{\xi_1+\ln a_{12}}{2}\bigg),\\
&S_2\,(\xi_2+\ln a_{12}\approx 0,\,\xi_1\to +\infty):\,f\sim 1+a_{12}\e^{\xi_2},\,u\sim \widehat{u_{2}}=\frac{k_2^2}{2}\sech^2\bigg(\frac{\xi_2+\ln a_{12}}{2}\bigg);
\end{split}\label{xasy02}
\end{flalign}
The constant length stem:
\begin{flalign}
	\begin{split}
&S_{1-2}\, (\xi_1\approx \xi_2,\,\xi_{1,2}\to +\infty):\,f\sim \e^{\xi_1}+\e^{\xi_2},\,u\sim u_{1-2}=\frac{(k_1- k_2)^2}{2}\sech^2\bigg(\frac{\xi_1- \xi_2}{2}\bigg).
	\end{split}\label{xstem01}
\end{flalign}

\begin{rk}
In section \ref{sec2}, $S_j$ corresponds to the formula $u_j$, $\widehat{u_j}$ and $\widetilde{u_j}$. 
The difference between $u_j$ and $\widehat{u_j}$ (or $\widetilde{u_j}$) is that the former does not contain $a_{12}$ while the latter does. 
So does $l_j$ and $\widehat{l_j}$ (or $\widetilde{l_j}$). 
The trajectories of $S_j$ before collision are $l_j$, the analogue after collision are $\hat{l}_j$ and the location interior stem is described by $l_{1-2}$. 
All of them are plotted in Fig. \ref{fig2s-1} (b).
\end{rk}
\begin{table}
  \centering
  \begin{tabular}{cccccc}
    \Xhline{1pt}
    Arm & Velocity ($(x,\,y)$-direction) & Amplitude  & Trajectory &Component \\
    \hline
    \multirow{3}{*}{$S_j$} & \multirow{3}{*}{$(k_j^2+\frac{3p_j^2}{k_j^2}, \,\frac{k_j^4+3p_j^2}{k_jp_j})$} & \multirow{3}{*}{$\frac{k_j^2}{2}$ }  &$l_j$  & $u_{j}$ \\
       &  & & $\widehat{l_j}$ & $\widehat{u_{j}}$ \\
       &  & & $\widetilde{l_j}$ & $\widetilde{u_{j}}$ \\
     \Xhline{1pt}
   $S_{1-2}$  & $(v^{1-2}_{[x]},\,v^{1-2}_{[y]})$ & $\frac{(k_1-k_2)^2}{2}$ & $l_{1-2}$& $u_{1-2}$ \\
     \hline
   $S_{1+2}$ & $(v^{1+2}_{[x]},\,v^{1+2}_{[y]})$ & $\frac{(k_1+k_2)^2}{2}$ & $l_{1+2}$ & $u_{1+2}$ \\
     \Xhline{1pt}
  \end{tabular}
  \caption{Physical quantities of the arms in section \ref{sec2} (KPII equation). 
  The arms $S_j$ $(j=1,2,1\pm2)$ correspond to $u_j$ ,\,$\widehat{u_j}$ or $\widetilde{u_j}$. 
  The relevant formulas are listed by Eqs.\ \eqref{xasy01}--\eqref{v01} and \eqref{xasy04}--\eqref{v03}.}\label{tab:t1}
\end{table}

Table \ref{tab:t1} provides the formulas, trajectories, amplitudes, and velocities for each arm, where
\begin{equation}\label{l01}
\boldsymbol{l_j:}\,\xi_j=0,\,\quad \boldsymbol{\widehat{l_j}:}\,\xi_j+\ln a_{12}=0,\,\quad \boldsymbol{l_{1-2}:}\, \xi_1-\xi_2=0,\,\quad j=1,\,2.
\end{equation}
and
\begin{equation}\label{v01}
v^{1-2}_{[x]}=k_1^2+k_1k_2+k_2^2+\frac{3p_1^2k_2-3p_2^2k_1}{k_1k_2(k_1-k_2)},\,v^{1-2}_{[y]}=\frac{k_1k_2(k_1^3-k_2^3)-3k_1p_2^2+3k_2p_1^2}{k_1k_2(p_1-p_2)}.
\end{equation}

Specifically, the trajectories of the arms are shown in Fig. \ref{fig2s-1} (b). 
According to Table \ref{tab:t1}, the amplitude of the stem \( S_{1-2} \) is larger than the arms \( S_1 \) and \( S_2 \) when \( k_2(k_2-2k_1) > 0 \) and \( k_1(k_1-2k_2) > 0 \). 
Conversely, the amplitude of the stem \( S_{1-2} \) is less than that of the arms \( S_1 \) and \( S_2 \) when \( k_2(k_2-2k_1) < 0 \) and \( k_1(k_1-2k_2) < 0 \).

The stem \( S_{1-2} \), also referred to as a virtual soliton, was initially introduced in Ref.\ \cite{jpsj1980} for the extended Boussinesq-like equation and later in Ref.\ \cite{kpjpsj1983} for the Kadomtsev-Petviashvili equation. 
Recently, the localized characteristics of the stem structures in the ANNV system have been analyzed in Ref.\ \cite{yuan2024}. 
In this section, we employ a similar method to analyze the stem structure in the quasi-resonant 2-soliton solution of the KPII equation.

Solving a group of equations $\xi_1=0$ and $\xi_2=0$ implies an intersection point $A_1$ of $l_1$ and $l_2$ as on ($x,y$)-plane:
\begin{flalign}
\begin{split}\label{eqa1}
A_1:\, \left(v^{A}_{[x]}t,\,v^{A}_{[y]}t \right),
\end{split}
\end{flalign}
where
\begin{equation}\label{v02}
v^{A}_{[x]}=\frac{k_1p_2^3-k_2^3p_1}{k_1p_2-k_2p_1}-\frac{3k_1k_2}{p_1p_2},\,v^{A}_{[y]}=-\frac{k_1k_2(k_1^2-k_2^2)}{k_1p_2-k_2p_1}+\frac{3(k_1p_2+k_2p_1)}{k_1k_2}.
\end{equation}

Similarly, by solving a group of $\xi_1+\ln a_{12}=0$ and $\xi_2+\ln a_{12}=0$, the intersection point $B_1$ on ($x,y$)-plan of $\widehat{l_1}$ and $\widehat{l_2}$ can be generated as
\begin{flalign}
\begin{split}\label{eqb1}
B_1:\, \left(\frac{(p_1-p_2)\ln a_{12}}{k_1p_2-k_2p_1}+v^{A}_{[x]}t,\,-\frac{(k_1-k_2)\ln a_{12}}{k_1p_2-k_2p_1}+v^{A}_{[y]}t\right).
\end{split}
\end{flalign}
It is noteworthy that points $A_1$ and $B_1$ also serve as the endpoints of $l_{1-2}$, which can be seen in Fig. \ref{fig2s-1} (b). Consequently, the length of the stem, denoted as $L_{A_1B_1}$, is defined as:
\begin{equation}\label{eqab1}
L_{A_1B_1} = \left| \frac{\ln a_{12}}{k_1p_2-k_2p_1} \right| \sqrt{(k_1-k_2)^2+(p_1-p_2)^2}.
\end{equation}

\begin{figure}[h!tb]
	\centering
    \subfigure[]{\includegraphics[height=4cm,width=4cm]{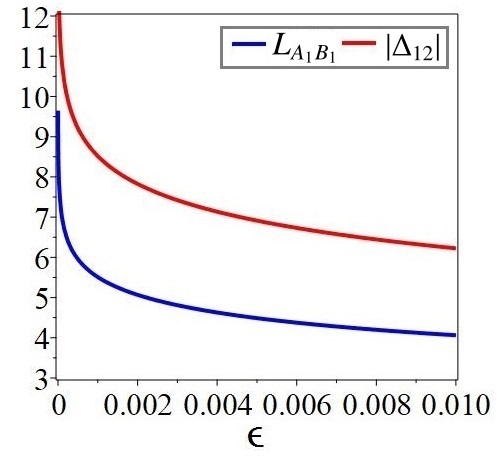}}
    \subfigure[]{\includegraphics[height=4cm,width=5.5cm]{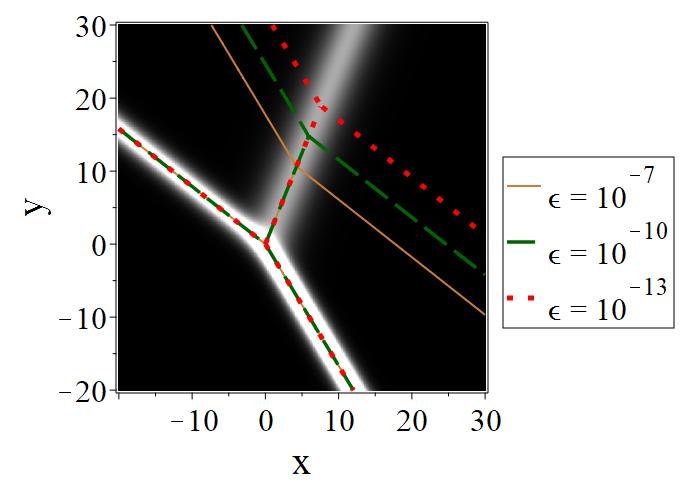}}
    \subfigure[]{\includegraphics[height=4cm,width=7cm]{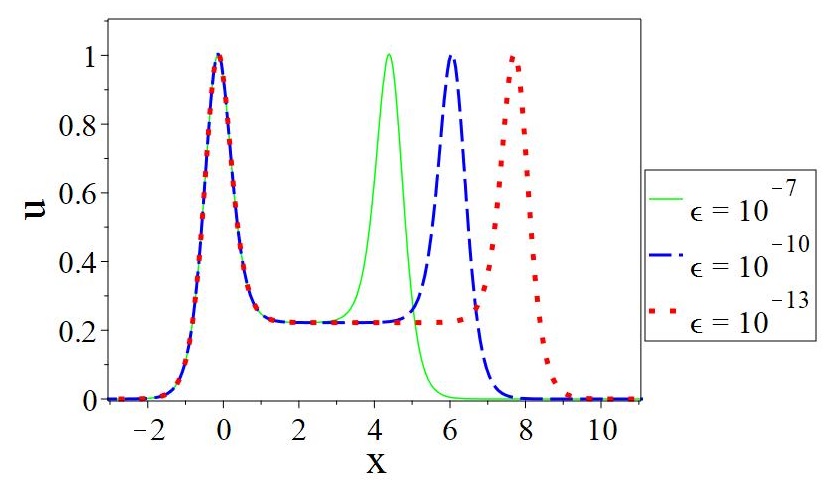}}
	\caption{Parameters: $k_1=\frac{5}{3},\,k_2=1,\,k_3=1,\,p_1=1,\,p_2=\frac { k_2(k_1^{2}-k_1k_2+ p_1)}{k_1}-\epsilon
,\,t=0$. (a) Graphs of \( L_{A_1B_1} \) and \(|\Delta_{12}| \) as the function of \( \epsilon \); (b) The trajectories of $u_{qw}^{[2]}$ with different $\epsilon$; (c) The section curves $u|_{l_{1-2}}$ with different $\epsilon$.}\label{fig2s-2}
\end{figure}

Fig. \ref{fig2s-2} (a) shows the trend of \( L_{A_1B_1} \) and the phase shift \(|\Delta_{12}| \) over \( \epsilon \) where $\epsilon$ comes from the choices
to implement weakly quasi-collision condition, see formulas of $P_2$ at the beginning of section \ref{sec2.1}. 
It can be confirmed from both the formulas and the figures that the smaller \( \epsilon \) is, the larger \( L_{A_1B_1} \) and \( |\Delta_{12}| \) are. 
When $\epsilon\to 0$, we have $a_{12}\to 0$ and $L_{A_1B_1} \to+\infty$, then the 2-soliton becomes to weakly resonant soliton (Y-shaped soliton). 
Fig. \ref{fig2s-2} (a) shows the trajectories of the weakly quasi-resonant solitons with different \( \epsilon \), where the background plane is a density map of the weakly resonant 2-soliton solution \eqref{weakresonance}. 
As can be seen from the figure, there is a pair of V-shaped solitons connected by a central stem structure in both the northeast and southwest directions. 
As the value of \( \epsilon \) decreases, the V-shaped soliton in the northeast moves farther away from the V-shaped soliton in the southwest, causing the stem structure to become longer. 
Until $\epsilon =0$, the stem structure becomes infinitely long, then the quasi-resonant soliton turns to the resonant Y-shaped soliton corresponding to the background plane. 
Both of these subgraphs confirm that the weakly resonant soliton is the limit state of the weakly quasi-resonant soliton, and the weakly quasi-resonant soliton is the intermediate state between the X-shaped soliton and the Y-shaped soliton.

Next, we study the cross-sectional curve of the stem structure $S_{1-2}$. 
The cross-sectional curve of the 2-soliton $u_{qw}^{[2]}$ on the planes $\xi_1-\xi_2=0$ is explored as
\begin{flalign}\label{cross2s01}
\begin{split}
u|_{l_{1-2}}=\frac{2\bigg((k_1+k_2)^2a_{12}+(k_1-k_2)^2+(k_1^2+k_2^2)(a_{12}\e^{\Theta_1}+\e^{-\Theta_1})\bigg)}{(a_{12}\e^{\Theta_1}+\e^{-\Theta_1}+2)^2},
\end{split}
\end{flalign}
where $\Theta_1=-\frac{(k_1p_2-k_2p_1)x}{p_1-p_2}+\bigg(\frac{k_1^3p_2-k_2^3p_1}{p_1-p_2}-\frac{3p_1p_2(k_1p_2-k_2p_1)}{k_1k_2(p_1-p_2)}\bigg)t$. 
The cross-sectional curves with different $\epsilon$ are illustrated in Fig. \ref{fig2s-2} (c), which also confirms that the smaller $\epsilon$ is, the longer the stem is. 
Deriving the extreme values by taking the derivative of Eq.\ \eqref{cross2s01}, we observe that instead of a line soliton having an extreme value line, the stem structure $S_{1-2}$ possesses only one extreme point between $A_1$ and $B_1$, denoted as $M_1$. 
The extreme point $M_1$ has the coordinate on the $(x, y)$-direction as follows:
\begin{flalign}\label{extremr1}
\begin{split}
M_1:\,\left(\frac{(p_1-p_2)\ln a_{12}}{2(k_1p_2-k_2p_1)}+v^{A}_{[x]}t,\,-\frac{(k_1-k_2)\ln a_{12}}{2(k_1p_2-k_2p_1)}+v^{A}_{[y]}t\right)
\end{split}
\end{flalign}
It is noteworthy that \( M_1 \) precisely corresponds to the midpoint of \( A_1B_1 \), illustrated in Fig. \ref{fig2s-1} (c).

Substituting \eqref{extremr1} into \eqref{cross2s01}, we obtain the extreme values of \( S_{1-2} \) as

\[
u(M_1) = \frac{k_1^2 + k_2^2}{2} + \frac{k_1k_2(a_{12}-1)}{(1 + \sqrt{a_{12}})^2}.
\]
Because of $\underset{\epsilon \to 0} \lim\,a_{12} =0$, it is easy to get $\underset{\epsilon \to 0} \lim\,u(M_1) = \frac{(k_1 - k_2)^2}{2}$. 
Therefore, when \( \epsilon \approx 0 \), we have \( u(M_1) \approx \frac{(k_1 - k_2)^2}{2} \). 
We can see that constant height of virtual soliton $S_{1-2}$ is $\frac{(k_1-k_2)^2}{2}$, which is same as the limit of $u(M_1)$. 
Thus the stem structure is described exactly by $u|_{l_{1-2}}$ in Eq.\ \eqref{cross2s01}, but the virtual soliton can provide an excellent  approximation of the bottom (almost flat) part the stem. 
This statement is confirmed by Figs. \ref{fig2s-1} (c) and \ref{fig2s-2} (c).

\subsection{Stem structure in strongly quasi-resonant soliton}\label{sec2.2}
In situations where $a_{12}\approx \infty,(\Delta_{12}\approx+\infty)$, the 2-soliton undergoes strongly quasi-resonant collisions. 
Making the transformation $\xi_1\to\xi_1-\ln a_{12}$, tau function \eqref{2f} becomes
\begin{flalign}
	\begin{split}
f_{qs}^{[2]}=1+\frac{1}{a_{12}}\e^{\xi_1}+\e^{\xi_2}+\e^{\xi_1+\xi_2}.
	\end{split}\label{2f2}
\end{flalign}
And then the strongly quasi-resonant 2-soliton is given by $u_{qs}^{[2]}=2(\ln f_{qs}^{[2]})_{xx}$.
\begin{rk}
Strongly quasi-resonant solitons can also be obtained by directly using the tau function \eqref{2f} and taking $a_{12}\approx \infty$ (see the Refs.\ \cite{kpjpsj1983,yuan2024}). 
The transformation $\xi_1\to\xi_1-\ln a_{12}$ is done here to be consistent with the strongly resonant soliton \eqref{strongresonance}.
\end{rk}

To ensure $a_{12}\approx +\infty$, we must choose $p_2=-\frac { k_2(k_1^{2}+k_1k_2- p_1)}{k_1}-\epsilon$ or $p_2=\frac { k_2(k_1^{2}+k_1k_2+ p_1)}{k_1}+\epsilon$. Substituting them into $a_{12}$, we have $a_{12}=1+\frac{4k_1k_2^3}{2\epsilon k_2(k_1+k_2)+\epsilon^2}$. 
Not hard to find out if $k_1=-k_2$ then $a_{12}<0$. In order to ensure the smoothness of the solution, there are following five cases: (1) $k_1+k_2>0,\,k_1>0,\,\epsilon>0$; (2) $k_2+k_1>0,\,k_1<0,\,k_2>0,\,\epsilon<0$; (3) $k_1+k_2<0,\,k_1<0,\,k_2<0,\,\epsilon>0$; (4) $k_1+k_2<0,\,k_1>0,\,k_2<0,\,\epsilon<0$; (5) $k_2+k_1<0,\,k_1<0,\,k_2>0,\,\epsilon>0$. 
Without loss of generality, we just consider the case where $p_2=-\frac { k_2(k_1^{2}+k_1k_2- p_1)}{k_1}-\epsilon,\,k_1>0,\,k_2>0,\,\epsilon>0$.

\begin{figure}[h!tb]
	\centering
    \subfigure[3D plot]{\includegraphics[height=4cm,width=4cm]{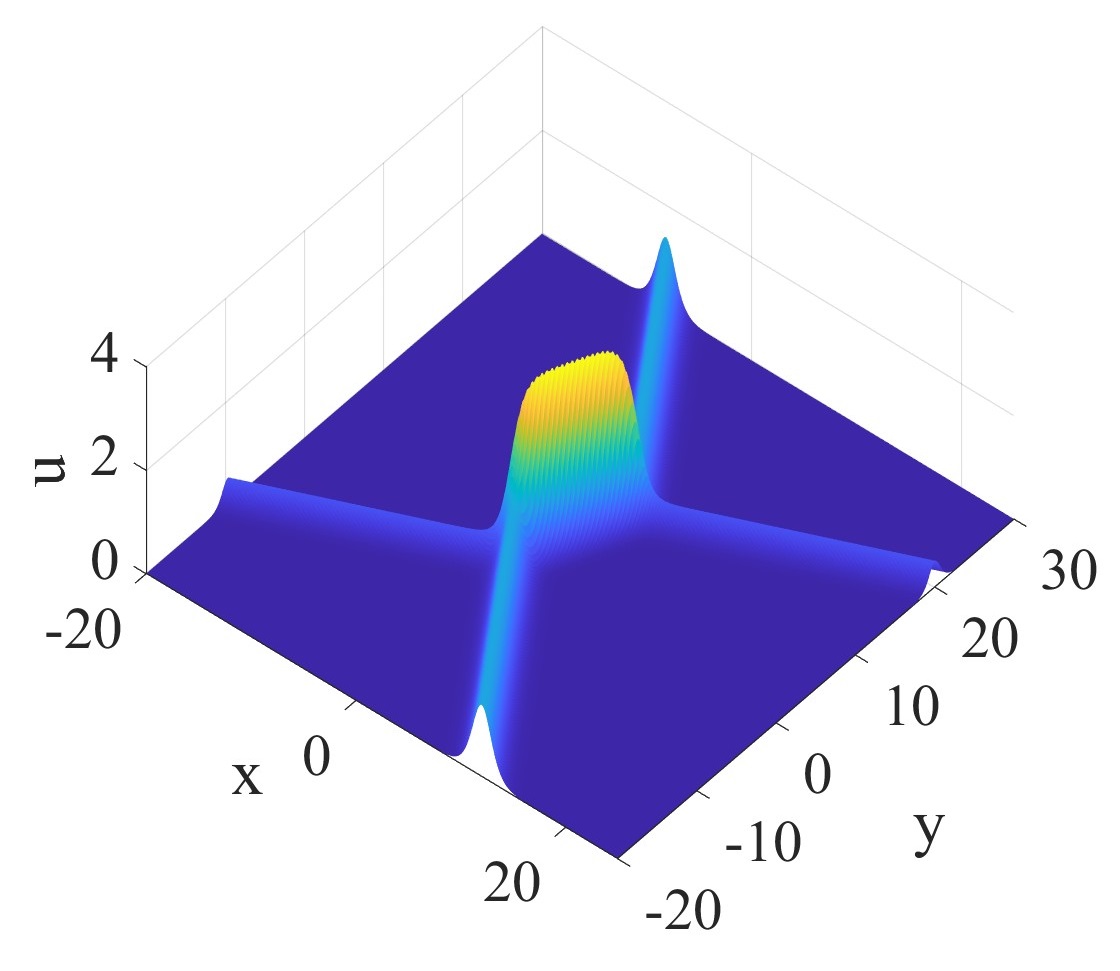}}
    \subfigure[Density map]{\includegraphics[height=4cm,width=5cm]{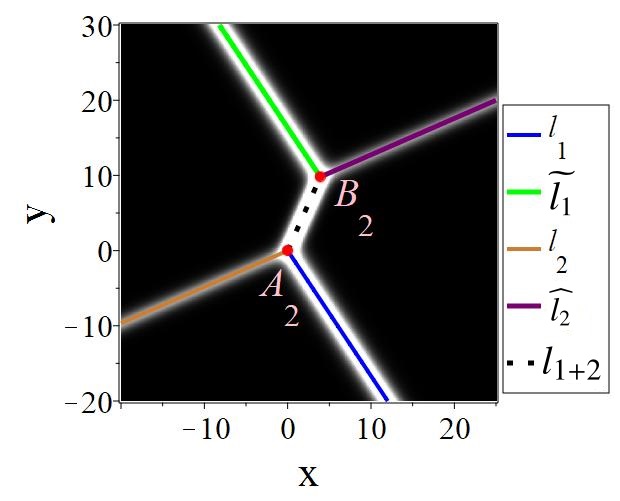}}
    \subfigure[Section curve $u|_{l_{1+2}}$]{\includegraphics[height=4cm,width=4cm]{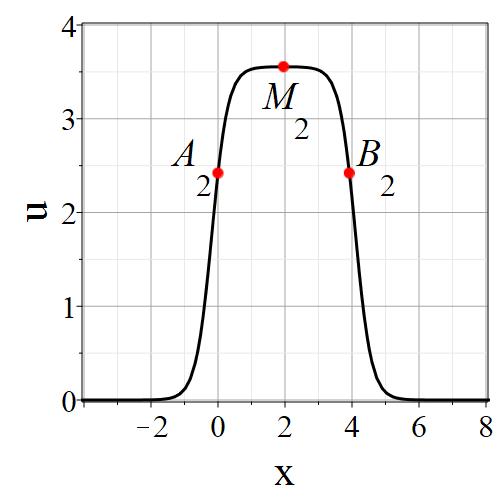}}
	\caption{The strongly quasi-resonant 2-soliton with $k_1=\frac{5}{3},\,k_2=1,\,p_1=1,\,p_2=-\frac {k_2(k_1^{2}+k_1k_2-p_1)}{k_1}-\epsilon,\,\epsilon=10^{-7}
,\,t=0$. (a) 3D map; (b) The density plot and trajectories; (c) The section-cross curve $u|_{l_{1+2}}$}\label{fig2s-3}
\end{figure}

In analogy to the weakly quasi-resonant 2-soliton, the strongly quasi-resonant 2-soliton, shown in Fig.\ \ref{fig2s-3} (a), exhibits the following asymptotic forms:\\
\begin{flalign}
\begin{split}
&S_1\,(\xi_1\approx 0,\,\xi_2\to -\infty):\,f\sim 1+\e^{\xi_1},\,u\sim u_{1}=\frac{k_1^2}{2}\sech^2\left(\frac{\xi_1}{2}\right),\\
&S_2\,(\xi_2\approx 0,\,\xi_1\to -\infty):\,f\sim 1+\e^{\xi_2},\,u\sim u_{2}=\frac{k_2^2}{2}\sech^2\left(\frac{\xi_2}{2}\right),\\
\end{split}\label{xasy03}
\end{flalign}
After collision:
\begin{flalign}
\begin{split}
&S_1\, (\xi_1-\ln a_{12}\approx 0,\,\xi_2\to +\infty):\,f\sim 1+\frac{1}{a_{12}}\e^{\xi_1},\,u\sim \widetilde{u_{1}}=\frac{k_1^2}{2}\sech^2\left(\frac{\xi_1-\ln a_{12}}{2}\right),\\
&S_2\,(\xi_2+\ln a_{12}\approx 0,\,\xi_1\to +\infty):\,f\sim 1+a_{12}\e^{\xi_2},\,u\sim \widehat{u_{2}}=\frac{k_2^2}{2}\sech^2\left(\frac{\xi_2+\ln a_{12}}{2}\right),
\end{split}\label{xasy04}
\end{flalign}
The constant length stem:
\begin{flalign}
	\begin{split}
&S_{1+2}\, (\xi_1\approx -\xi_2,\,\xi_1\to +\infty,\,\xi_2\to -\infty):\,f\sim 1+\e^{\xi_1+\xi_2},\,u\sim u_{1+2}=\frac{(k_1+ k_2)^2}{2}\sech^2\left(\frac{\xi_1+ \xi_2}{2}\right).
	\end{split}\label{xstem02}
\end{flalign}

The trajectories, amplitudes, velocities of these five arms before and after collision are provided in table \ref{tab:t1} and trajectories are plotted in Fig. \ref{fig2s-3} (b), where
\begin{equation}\label{l02}
\boldsymbol{\widetilde{l_{1}}:}\quad \xi_1-\ln a_{12}=0,\qquad\boldsymbol{l_{1+2}:}\quad \xi_1+\xi_2=0.
\end{equation}
and
\begin{equation}\label{v03}
v^{1+2}_{[x]}=k_1^2-k_1k_2+k_2^2+\frac{3p_1^2k_2+3p_2^2k_1}{k_1k_2(k_1+k_2)},\,v^{1+2}_{[y]}=\frac{k_1k_2(k_1^3+k_2^3)+3k_1p_2^2+3k_2p_1^2}{k_1k_2(p_1+p_2)}.
\end{equation}

Now we will figure out the coordinates of the two endpoints (noted as $A_2$ and $B_2$) and length of the stem structure. 
Solving the system of equations $\xi_1=0$ and $\xi_2=0$ leads to an intersection point of $l_1$ and $\widehat{l_2}$:
\begin{flalign}
\begin{split}\label{eqa2}
A_2:\, \left(v^{A}_{[x]}t,\,v^{A}_{[y]}t \right).
\end{split}
\end{flalign}
Similarly, the intersection point of $\widetilde{l_1}$ and $l_2$ can be expressed by:
\begin{flalign}
\begin{split}\label{eqb2}
B_2:\, \left(\frac{(p_1+p_2)\ln a_{12}}{k_1p_2-k_2p_1}+v^{A}_{[x]}t,\,-\frac{(k_1+k_2)\ln a_{12}}{k_1p_2-k_2p_1}+v^{A}_{[y]}t \right).
\end{split}
\end{flalign}
The trajectories of arms and the two endpoints of the stem are shown in Fig.\ \ref{fig2s-3} (b). 
Then the length of the stem $S_{1+2}$ can be obtained as
\begin{equation}\label{eqab2}
L_{A_2B_2}=\left| \frac{\ln a_{12}}{k_1p_2-k_2p_1} \right|\sqrt{(k_1+k_2)^2+(p_1+p_2)^2}.
\end{equation}

\begin{figure}[h!tb]
	\centering
  \subfigure[]{\includegraphics[height=4cm,width=4cm]{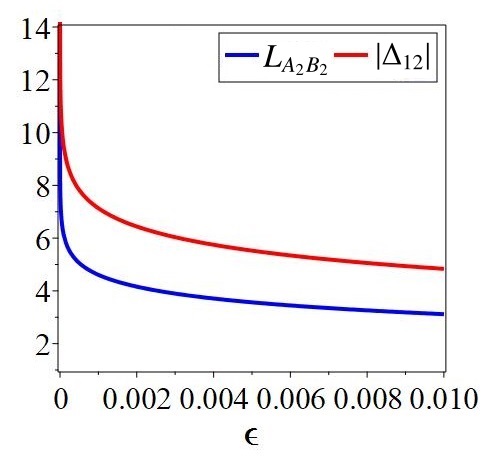}}
  \subfigure[]{\includegraphics[height=4cm,width=5.5cm]{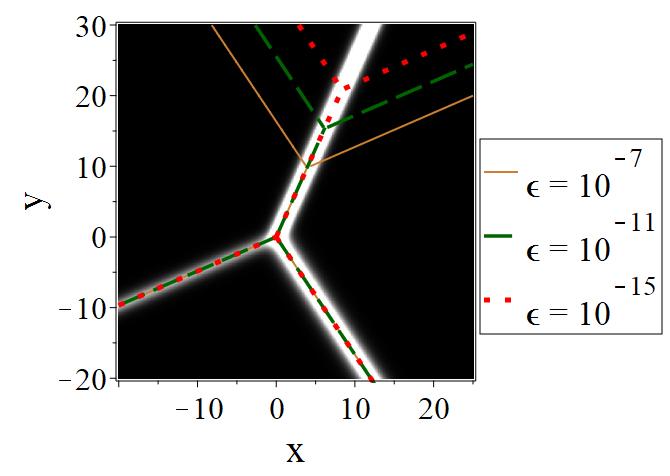}}
 \subfigure[]{\includegraphics[height=4cm,width=6cm]{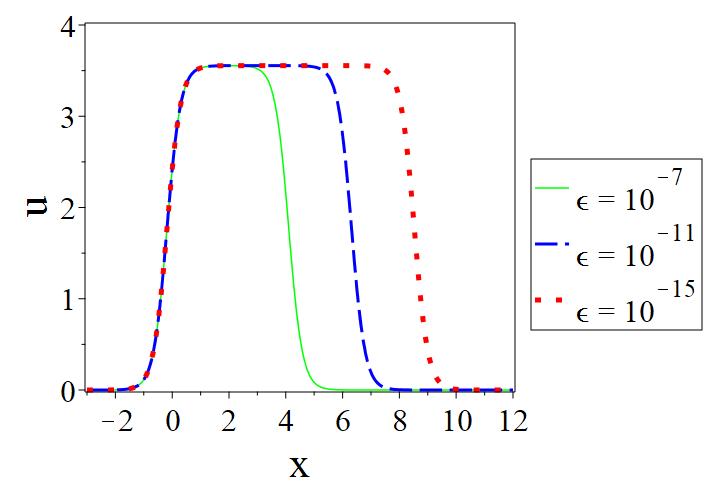}}
	\caption{Parameters: $k_1=\frac{5}{3},\,k_2=1,\,p_1=1,\,p_2=-\frac {k_2(k_1^{2}+k_1k_2-p_1)}{k_1}-\epsilon
,\,t=0$. (a) Graphs of \( L_{A_2B_2} \) and \(|\Delta_{12}| \) as the function of \( \epsilon \); 
(b) The trajectories of $u_{qs}^{[2]}$ with different $\epsilon$; (c) The section curves $u|_{l_{1+2}}$ with different $\epsilon$.}\label{fig2s-4}
\end{figure}

Fig.\ \ref{fig2s-4} (a) illustrates the behavior of \( L_{A_2B_2} \) and \( |\Delta_{12}| \) as functions of \(\epsilon\) where $\epsilon$ comes from the choices
to implement strongly quasi-collision condition, see formulas of $P_2$ at the beginning of section \ref{sec2.2}. 
The equations and the figure demonstrate that as \(\epsilon\) decreases, both \( L_{A_2B_2} \) and \( |\Delta_{12}|\) increase. 
In the limit as \(\epsilon \to 0\), \( a_{12} \to +\infty \) and \( L_{A_2B_2} \to +\infty \), resulting in the transformation of the 2-soliton into a strongly resonant soliton, also referred to as a Y-shaped soliton. 
On the other hand, the trajectories of the strongly quasi-resonance solitons for varying \(\epsilon\) values are depicted in Fig.\ \ref{fig2s-4} (b), where the background plane is the density map of the strongly resonant 2-soliton solution given by Eq.\ \eqref{strongresonance}. 
It shows that two pairs of V-shaped solitons connected by a central stem structure oriented in both the northeast and southwest directions. 
As \(\epsilon\) decreases, the V-shaped solitons in the northeast and southwest move further apart, causing the central stem structure to elongate. 
Until \(\epsilon = 0\), the stem structure extends infinitely, and the quasi-resonant soliton transitions into the resonant soliton corresponding to the background plane. 
Thus, the strongly resonant soliton represents the limiting state of the strongly quasi-resonant soliton, which itself is an intermediate state between the X-shaped and Y-shaped solitons.

Upon comparison, we observe that Eqs.\ \eqref{eqab1} and \eqref{eqab2} are identical to formulas (3) and (4) in Ref.\ \cite{yuan2024}. 
Consequently, the trajectories of the four arms and the stem structure in different soliton equations are determined by the coefficients of \(x\) and \(y\) and $a_{12}$ in the tau function. 
This leads us to the following Remark:
\begin{rk}
If the 2-soliton of a soliton equation has a tau function given by \( f = 1 + e^{\xi_1} + e^{\xi_2} + a_{12}e^{\xi_1+\xi_2} \) with \( \xi_j = k_j x + p_j y + \omega_j t + \xi_j^0 \), the length of the stem structure in the quasi-resonant 2-soliton is respectively given by \eqref{eqab1} and \eqref{eqab2} with weakly and strongly case, where $k_j,\,p_j$ make sure $a_{12}\approx 0$ or $a_{12}\approx+\infty$ respectively. 
This means two formulas are universal for the above-mentioned tau function $f$.
\end{rk}

Subsequently, we study the cross-sectional curve of the stem $S_{1+2}$. 
The cross-sectional curve, situated on planes defined by $\xi_1+\xi_2=0$ of the 2-soliton $u_{qs}^{[2]}$, is shown in Fig. \ref{fig2s-3} (c) and formulated as follows:
\begin{flalign}\label{cross2s02}
\begin{split}
u|_{l_{1+2}}=\frac{2a_{12}\bigg((k_1+k_2)^2a_{12}+(k_1-k_2)^2+(k_1^2+k_2^2)(\e^{\Theta_2}+a_{12}\e^{-\Theta_2})\bigg)}{(\e^{\Theta_2}+a_{12}\e^{-\Theta_2}+2a_{12})^2},
\end{split}
\end{flalign}
where $\Theta_2=\frac{(k_1p_2-k_2p_1)x}{p_1+p_2}-\bigg(\frac{k_1^3p_2-k_2^3p_1)}{p_1+p_2}-\frac{3p_1p_2(k_1p_2-k_2p_1)}{k_1k_2(p_1+p_2)}\bigg)t$. 
The cross-sectional curves for various \(\epsilon\) values are depicted in Fig.\ \ref{fig2s-4} (c), demonstrating that a smaller \(\epsilon\) results in a longer stem. 
By taking the derivative of Eq.\ \eqref{cross2s02} with respect to $x$ or $y$, we can identify their extreme values. 
Similar to \( S_{1-2} \), the stem \( S_{1+2} \) features only one extreme point between \( A_2 \) and \( B_2 \), and its coordinates on the \((x,y)\)-plane are given by
\begin{flalign}\label{extremr2}
\begin{split}
M_2:\,\left(\frac{(p_1+p_2)\ln a_{12}}{2(k_1p_2-k_2p_1)}+v^{A}_{[x]}t,\, -\frac{(k_1+k_2)\ln a_{12}}{2(k_1p_2-k_2p_1)}+v^{A}_{[y]}t\right).
\end{split}
\end{flalign}
In the same way, $M_2$ also precisely corresponds to the midpoint of $A_2B_2$, depicted in Fig. \ref{fig2s-3} (c).

Substituting \eqref{extremr2} into \eqref{cross2s02}, we obtain the extreme values of $S_{1+2}$ as
$$u(M_2)=\frac{(k_1+k_2)^2+\frac{(k_1-k_2)^2}{a_{12}}+\frac{k_1^2+k_2^2}{\sqrt{a_{12}}}}{2(1+\frac{1}{\sqrt{a_{12}}})^2}.$$
Because of $\underset{\epsilon \to 0}  \lim\,\frac{1}{a_{12}} =0$, it is easy to get $\underset{\epsilon\to 0} \lim\, u(M_2)=\frac{(k_1+k_2)^2}{2}$. 
That is, when $\epsilon\approx 0$, we have $u(M_2)\approx \frac{(k_1+k_2)^2}{2}$. 
We can see that constant height of virtual soliton $S_{1+2}$ is $\frac{(k_1+k_2)^2}{2}$, which is same as the limit of $u(M_2)$. 
Thus the stem structure is described exactly by $u|_{l_{1+2}}$ in Eq. \eqref{cross2s02}, but the virtual soliton Eq. \eqref{xstem02} can provide an excellent approximation of the top (almost flat) part the stem. 
This statement is confirmed by Figs. \ref{fig2s-3} (c) and \ref{fig2s-4} (c).

\subsection{Application of the quasi-resonant 2-soliton solution}\label{sec2.3}
\begin{figure}[h!tb]
	\centering
\includegraphics[height=3.5cm,width=9cm]{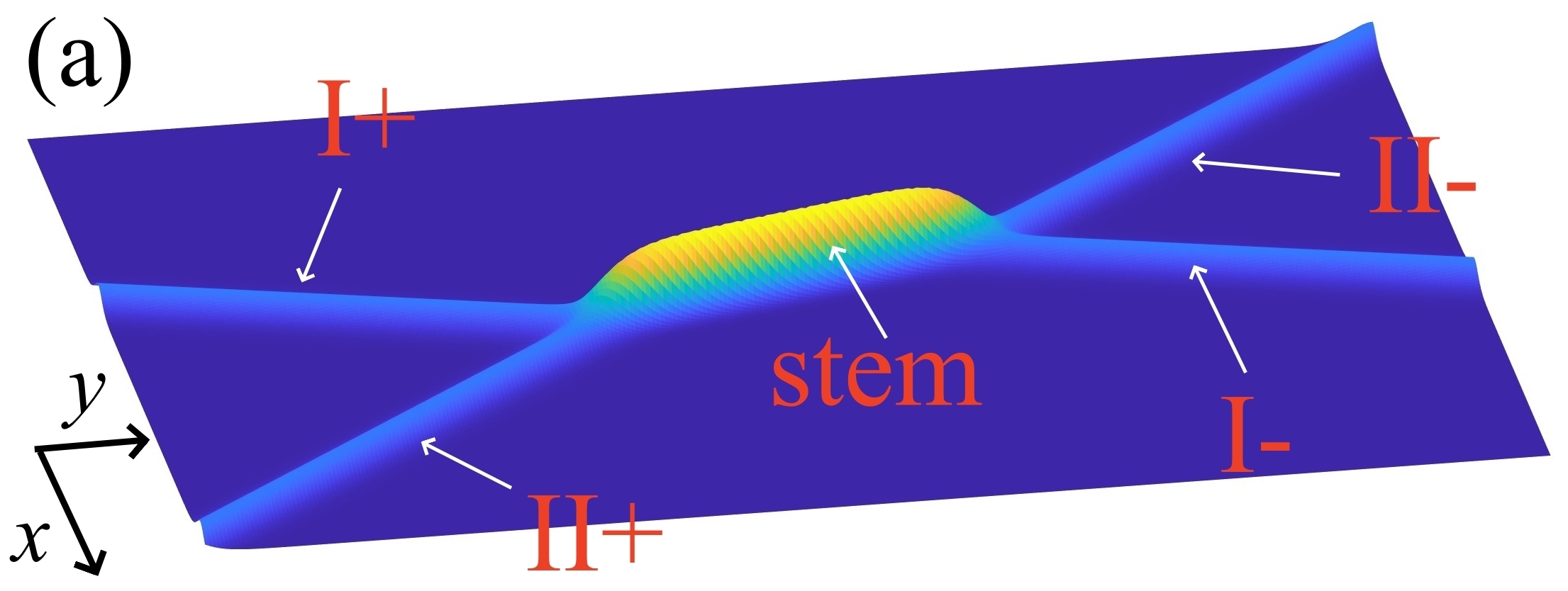}\\
 \quad\includegraphics[height=4cm,width=11.8cm]{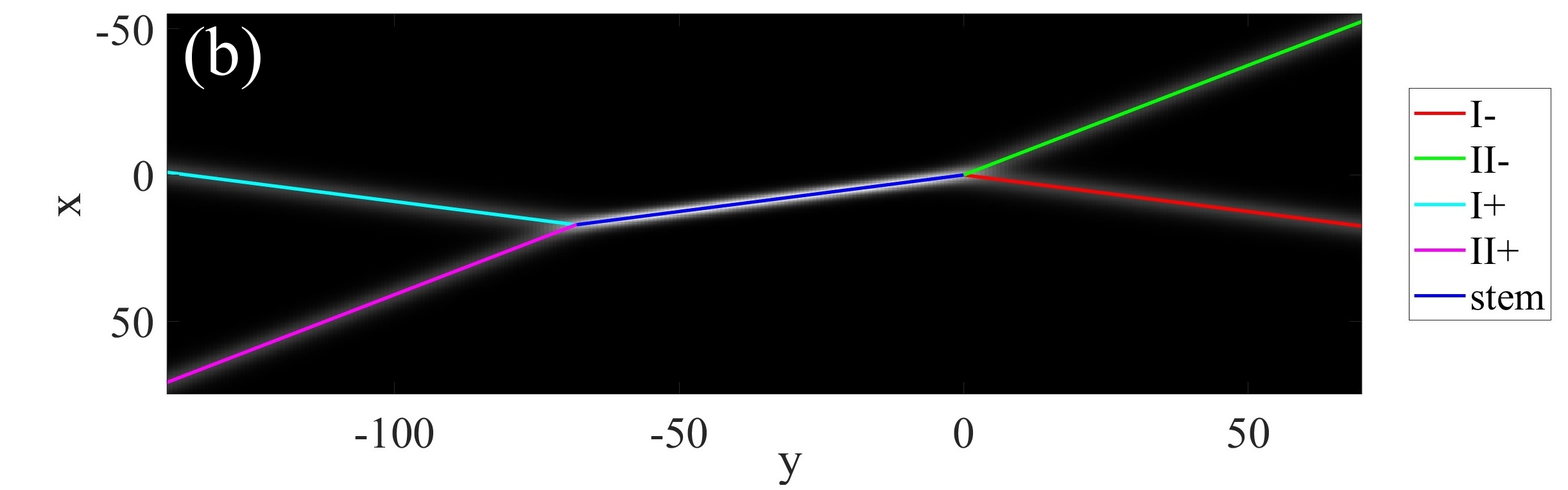}\\
 \includegraphics[height=3cm,width=9cm]{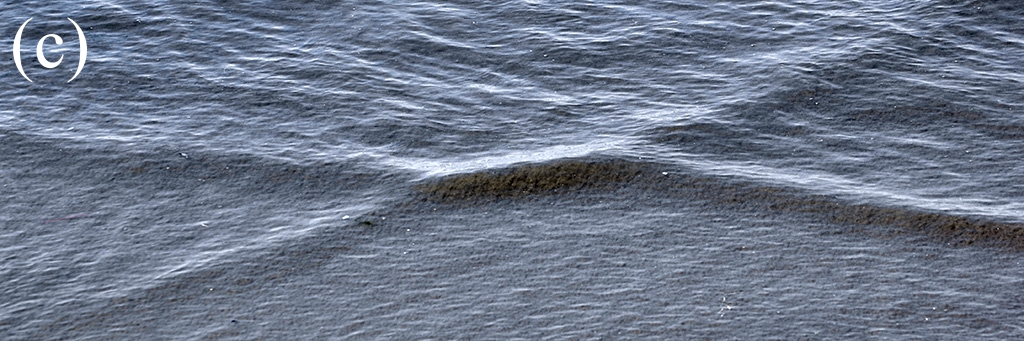}
	\caption{Comparison between plots(Fig.(a,b)) of quasi-resonant 2 soliton and photograph (c) for patterns of ocean wave. 
  The photograph originates from Figs. 3 (b) of Ref. \cite{kppre2012}. 
  Here, I and II denote soliton-1 and soliton-2, while $-$ and $+$ denote before and after collision, respectively. 
  Parameters: $k_1=k_2=\frac{1}{2},\,p_1=-\frac{1}{8}-10^{-8},\,p_2=\frac{3}{8},\,t=0$. }\label{fig2s-pre}
\end{figure}

In this subsection, we apply quasi-resonant 2-soliton solutions to model wave patterns observed in the ocean. 
Fig. \ref{fig2s-pre} (a) shows a photograph taken by Douglas Baldwin at Venice Beach, California \cite{kppre2012,photo,jpa2014}, in which the water waves exhibit two V-shaped profiles and a stem structure in between, corresponding to the quasi-resonant 2-soliton studied here. 
While the soliton arms and stem structures have been visualized through three-dimensional plots in Ref. \cite{kppre2012} (see Fig. 3) and briefly analyzed in Ref. \cite{jpa2014} (see Section 2.3), no detailed analysis of these features has been carried out. 
In this section, we will determine the precise locations of the observed arms and stem shown in Fig. 3(b) of Ref. \cite{kppre2012} (or Fig. 6(b) of Ref. \cite{jpa2014}) based on the quasi-resonant 2-soliton results.

In the case of $k_1=k_2=\frac{1}{2},\,p_1=-\frac{1}{8}-10^{-8},\,p_2=\frac{3}{8},\,t=0$ (corresponding to Fig. \ref{fig2s-pre} (b) and (d)), \( a_{12} \approx 2.50\times 10^8 \approx +\infty \) can be obtained, representing the strongly quasi-resonant scenario. 
The analytical expression is provided in \eqref{2f2} and \eqref{equ}. 
By applying the same method as in the previous subsections, referring  to Eqs. \eqref{l01} and \eqref{l02}, we can derive the trajectories of the four arms and the stem structure which are shown in Fig. \ref{fig2s-pre} (c) as follows:
\begin{flalign}
\begin{split}\label{III1}
&\text{I-:}\,{\frac {x}{2}}-{\frac {y}{8}}=0,\,\text{I+:}\,{\frac {x}{2}}-{\frac {y}{8}}-\ln  (2.5\times 10^7)=0,\\
&\text{II-:}\,{\frac {x}{2}}+{\frac {3\,y}{8}}=0,\,\text{II+:}\,{\frac {x}{2}}+{\frac {3\,y}{8}}+\ln (2.5\times 10^7)=0,\\
&\text{Stem:}\,x+{\frac {y}{4}}=0.
\end{split}
\end{flalign}

While Ref. \cite{kppre2012} explains Fig. \ref{fig2s-pre} (c) solely within the framework of strong quasi-resonance, Ref. \cite{jpa2014} does not distinguish between strong and weak quasi-resonance, treating both simply as resonance. 
In contrast, we adopt the strong quasi-resonance scenario to model the water waves depicted in Fig. \ref{fig2s-pre} (a) and (b). 
Although it is possible to specify the height of the stem, as demonstrated in Eq. \eqref{cross2s02}, Eq. \eqref{kpeq} represents a dimensionless KP equation, meaning the actual amplitudes of the solutions can be rescaled, making a discussion of their specific values less relevant. 
Therefore, we do not further address the amplitudes in this context. 
Additionally, the scenario corresponding to Figs. 9--13 in Ref. \cite{jpa2014} involves the investigation of 3-soliton resonances, which lies beyond the scope of this study.

\section{Stem structure in the quasi-resonant breather-soliton of the KPI equation}\label{sec3}
In this section, we analyze the stem structure in the quasi-resonant collision of the KPI equation \eqref{kpeq} with $\delta=-3$. 
In this scenario, it is impossible for $a_{ij}$ to be either zero or infinity if they are real, which prevents the formation of a resonant soliton for the KPI equation. 
Likewise, a quasi-resonant soliton solution cannot be directly achieved by approximating $a_{ij}$ as either 0 or infinity when they are real. 
Consequently, we focus on the quasi-resonant collision between a breather and a line soliton, using equations \eqref{equ} and \eqref{nf}, and explore the local structure within the quasi-resonant breather-soliton solution.

\subsection{Basic summary of the breather-soliton}\label{sec3.1}
By setting $N=3$ and $\delta=-3$ in Eq.\ \eqref{nf}, the tau function for the 3-soliton solution of the KPI equation is given by:
\begin{align}
f^{[3]}&=1+\exp\xi_1+\exp\xi_2+\exp\xi_3+a_{12}\exp(\xi_1+\xi_2)+a_{13}\exp(\xi_1+\xi_3)\nonumber\\
&+a_{23}\exp(\xi_2+\xi_3)+a_{12}a_{13}a_{23}\exp(\xi_1+\xi_2+\xi_3).\label{3f}
\end{align}

By substituting $k_1 = a_1 + b_1i = k_2^*$ and $p_1 = c_1 + d_1i = p_2^*$ (i.e., $\xi_1 = \xi_2^*$ and $a_{13} = a_{23}^*$) into the above formulas, we can obtain the hybrid solution consisting of a breather and a soliton, i.e. breather-soliton solution. 
In this case, $a_{ij}$ can be rewritten as follows:
\begin{flalign}\label{a123}
\begin{split}
&a_{12}=\frac{(a_1d_1 - b_1c_1)^2 + b_1^2(a_1^2 + b_1^2)^2}{(a_1d_1 - b_1c_1)^2 - a_1^2(a_1^2 + b_1^2)^2},\,a_{13}=\frac{(q_1+q_2i)(q_3+q_4i)}{(n_1+n_2i)(n_3+n_4i)},\,a_{23}=\frac{(q_1-q_2i)(q_3-q_4i)}{(n_1-n_2i)(n_3-n_4i)},
\end{split}
\end{flalign}
where
\begin{flalign}\label{eqqn1}
\begin{split}
&q_1=k_3a_1^2 - a_1k_3^2 - b_1^2k_3 + b_1p_3 - k_3d_1,q_2=2a_1b_1k_3 - b_1k_3^2 - a_1p_3 + k_3c_1,\\
&q_3=k_3a_1^2 - a_1k_3^2 - b_1^2k_3 - b_1p_3 + k_3d_1,q_4=2a_1b_1k_3 - b_1k_3^2+ a_1p_3- k_3c_1,\\
&n_1=k_3a_1^2 + a_1k_3^2 - b_1^2k_3 - b_1p_3 + k_3d_1,n_2=2a_1b_1k_3 + b_1k_3^2 + a_1p_3- k_3c_1,\\
&n_3=k_3a_1^2 + a_1k_3^2 - b_1^2k_3 + b_1p_3 - k_3d_1,n_4=2a_1b_1k_3 + b_1k_3^2- a_1p_3 + k_3c_1.
\end{split}
\end{flalign}
Then the tau function can be written as:
\begin{flalign}
f_{bs}=1+2\,{{\rm e}^{\theta_1}}\cos\eta_1 +{{\rm e}^{\xi_3}}+a_{12}{{\rm e}^{2\,\theta_1}}+2\,{{\rm e}^{\theta_1+\xi_3}}(
\alpha_1\cos\eta_1 -\beta_1\sin\eta_1) +a_{12}( \alpha_1^{2}+\beta_1^{2}) {{\rm e}^{2\,\theta_1+\xi_3}},\label{3fbs}
\end{flalign}
where
\begin{flalign}\label{alphabeta}
\begin{split}
&\alpha_1=(q_1q_3-q_2q_4)(n_1n_3 -n_2n_4)+ (q_1q_4+q_2q_3)(n_1n_4 +n_2n_3)=Re(a_{13}),\\
&\beta_1=(q_1q_4+q_2q_3)(n_1n_3 -n_2n_4)-(q_1q_3-q_2q_4)(n_1n_3+n_2n_4)=Im(a_{13}),\\
&\theta_1=a_1x+c_1y-(a_1^3-3a_1b_1^2)t+\frac{(3a_1c_1^2-3a_1d_1^2 +6b_1c_1d_1)t}{a_1^2 + b_1^2}=Re(\xi_1),\\
&\eta_1=b_1x+d_1y+(b_1^3 - 3a_1^2b_1)t+\frac{(3b_1d_1^2 - 3b_1c_1^2 + 6a_1c_1d_1)t}{a_1^2 + b_1^2}=Im(\xi_1),\\
&\xi_3=k_3x+p_3y+-\frac{k_3^4-3 p_3^2}{k_3}t,
\end{split}
\end{flalign}
and the smoothness condition for the breather-soliton $u_{bs}=2 (ln f_{bs})_{xx}$ is $a_{12} \geqslant 1$ and $\alpha_1^2 + \beta_1^2 \geqslant 0$. 
The breather-soliton $u_{bs}$ has the following asymptotic forms:

Before collision:

The breather $(\theta_1\approx 0,\,\xi_3\to -\infty)$:
\begin{flalign}
f\sim f_{B}^-= 1+2\e^{\theta_1}\cos\eta_1+a_{12}\e^{2\theta_1}.\label{xasy11}
\end{flalign}

The soliton $(\theta_1\to -\infty,\,\xi_3\approx 0)$:
\begin{flalign}
f\sim f_{S}^-= 1+\e^{\xi_3},\label{xasy12}
\end{flalign}

After collision:

The breather $(\theta_1\approx 0,\,\xi_3\to +\infty)$:
\begin{flalign}
f\sim f_{B}^+= 1+2\e^{\theta_1}(\alpha_1\cos\eta_1-\beta_1\sin\eta_1)+a_{12}(\alpha_1^2+\beta_1^2)\e^{2\theta_1},\label{xasy13}
\end{flalign}

The soliton $(\theta_1\to +\infty,\,\xi_3\approx 0)$:
\begin{flalign}
f\sim f_{S}^+= 1+(\alpha_1^2+\beta_1^2)\e^{\xi_3},\label{xasy14}
\end{flalign}

From the above asymptotic forms, we obtain the phase shifts of the breather ($\Delta_{B}$) and the soliton ($\Delta_{S}$) respectively as follows:
\begin{equation}\label{delta1}
\Delta_{B}=\frac{1}{2}\ln (\alpha_1^2+\beta_1^2),\,\Delta_{S}=\ln (\alpha_1^2+\beta_1^2).
\end{equation}

For convenience, let us denote $\Delta_{13} = \frac{1}{2} \ln (\alpha_1^2 + \beta_1^2)$. 
It is known that when the phase shift $\Delta_{13}$ is small, the oblique  breather-soliton solution exhibits an X-shape \cite{kp2019,kp2020}. 
Conversely, when the phase shift $\Delta_{13}$ becomes infinite, the oblique soliton and breather interact and merge to form a new soliton arm, resulting in a resonant solution with a Y-shape \cite{bs2021}. 
This raises the question: is there also a quasi-resonant collision between a breather and a soliton, similar to the two-soliton interaction studied in the previous section? 
The quasi-resonance and local structures (also referred to as stem structures) in the breather-soliton solution of the KPI equation are the primary focus of our research in this section. Analogous to the quasi-resonant two-soliton case, we define $\Delta_{13} \approx -\infty$ ($a_{13} \approx 0$) as weak quasi-resonance and $\Delta_{13} \approx +\infty$ ($a_{13} \approx +\infty$) as strong quasi-resonance. 
In the following, we will investigate these two cases separately.
\subsection{Stem structure in the weakly quasi-resonant breather-soliton}\label{sec3.2}
To derive the weakly quasi-resonant breather-soliton solution from $u_{bs}$ given in last subsection, it is necessary to first establish the parameter conditions required for resonance. Analysis of Eq.\ \eqref{a123} reveals that the resonance condition $\Delta_{13} = -\infty$ corresponds to either $q_1 = q_2 = 0$ or $q_3 = q_4 = 0$. 
Consequently, the resonance condition can be formulated as follows:
\begin{flalign}
	\begin{split}
k_3 = a_1-\frac{a_1 d_1-b_1 c_1}{a_1^2 + b_1^2}, \,
p_3 = a_1b_1+c_1-\frac{(a_1 d_1 - b_1 c_1) (a_1 c_1 + b_1 d_1)}{(a_1^2 + b_1^2)^2};
	\end{split}\label{kp01}
\end{flalign}
or
\begin{flalign}
	\begin{split}
k_3 = a_1+\frac{a_1 d_1-b_1 c_1}{a_1^2 + b_1^2}, \,
p_3 = -a_1b_1+c_1+\frac{(a_1 d_1 - b_1 c_1) (a_1 c_1 + b_1 d_1)}{(a_1^2 + b_1^2)^2}.
	\end{split}\label{kp02}
\end{flalign}

Therefore, the conditions that the parameters must satisfy for a weakly quasi-resonant collision are as follows:

(1) $k_3 = a_1-\frac{a_1 d_1-b_1 c_1}{a_1^2 + b_1^2}+\epsilon$, \quad (2) $p_3 =  a_1b_1+c_1-\frac{(a_1 d_1 - b_1 c_1) (a_1 c_1 + b_1 d_1)}{(a_1^2 + b_1^2)^2}+\epsilon$,

(3) $k_3 = a_1+\frac{a_1 d_1-b_1 c_1}{a_1^2 + b_1^2}+\epsilon$, \quad (4) $p_3 = -a_1b_1+c_1+\frac{(a_1 d_1 - b_1 c_1) (a_1 c_1 + b_1 d_1)}{(a_1^2 + b_1^2)^2}+\epsilon$.

Without loss of generality, we consider only the case (1) where \( k_3 = a_1 - \frac{a_1 d_1 - b_1 c_1}{a_1^2 + b_1^2} + \epsilon \) and \( p_3 = a_1 b_1 + c_1 - \frac{(a_1 d_1 - b_1 c_1)(a_1 c_1 + b_1 d_1)}{(a_1^2 + b_1^2)^2} \). 
Then the weakly quasi-resonant breather-soliton is given by
\begin{flalign}
&u_{b-s}=2(\ln f_{b-s})_{xx},\label{3u01}\\
&f_{b-s}=1+2\,{{\rm e}^{\theta_1}}\cos\eta_1 +{{\rm e}^{\xi_3}}+a_{12}{{\rm e}^{2\,\theta_1}}+2\,{{\rm e}^{\theta_1+\xi_3}}(
\alpha_1\cos\eta_1 -\beta_1\sin\eta_1) +a_{12}( \alpha_1^{2}+\beta_1^{2}) {{\rm e}^{2\,\theta_1+\xi_3}},\label{3f01}
\end{flalign}
where the relevant formulas are provided by Eqs.\ \eqref{eqqn1} and \eqref{alphabeta}. 
Using a similar approach to that of the quasi-resonant soliton, we can determine the asymptotic form of the intermediate stem structure as follows:
\begin{flalign}
\begin{split}
S_{1-3} \,(2\theta_1\approx \xi_3,\,\theta_1\to+\infty,\,\xi_3\to +\infty): \quad f\sim f_{1-3}= 1+a_{12}\e^{2\theta_1-\xi_3}.
\end{split}\label{xasy15}
\end{flalign}

Based on the above analysis, the weakly quasi-resonant breather-soliton \eqref{3u01} has the following asymptotic forms:\\
Before collision $(y\to-\infty)$:
\begin{flalign}
\begin{split}
&\mathbf{B_1}:\,\, f\sim 1+2\e^{\theta_1}\cos\eta_1+a_{12}\e^{2\theta_1},\,u\sim u_{B},\\
&\mathbf{S_3}:\,\, f\sim 1+\e^{\xi_3},\,u\sim u_{S}=\frac{k_3^2}{2}\sech(\xi_3);\\
\end{split}\label{asy11}
\end{flalign}
After collision $(y\to+\infty)$:
\begin{flalign}
\begin{split}
&\mathbf{B_1}:\,\, f\sim 1+2\e^{\theta_1}(\alpha_1\cos\eta_1-\beta_1\sin\eta_1)+a_{12}(\alpha_1^2+\beta_1^2)\e^{2\theta_1},\,u\sim \wideparen{u_{B}},\\
&\mathbf{S_3}:\,\, f\sim 1+(\alpha_1^2+\beta_1^2)\e^{\xi_3},\,u\sim \wideparen{u_{S}}=\frac{k_3^2}{2}\sech(\xi_3+\ln (\alpha_1^2+\beta_1^2));\\
\end{split}\label{asy12}
\end{flalign}
The stem structure:
\begin{flalign}
\begin{split}
\mathbf{S_{1-3}}:\,\, f\sim 1+a_{12}\e^{2\theta_1-\xi_3},\,u\sim u_{1-3}=\frac{(2a_1-k_3)^2}{2}\sech(2\theta_1-\xi_3+\ln a_{12}).\label{asystem1}
\end{split}
\end{flalign}

Here,
\begin{flalign}\label{uj03}
\begin{split}
u_{B}=&\frac{2\e^{\theta_1}\left(a_{12}r_4\e^{2\theta_1}+(4a_{12}a_1^2-b_1^2)\e^{\theta_1}+r_3 \right)}{(1 +\e^{\theta_1}\cos\eta_1 + a_{12}\e^{2\theta_1})^2},\\
\wideparen{u_{B}}=&\frac{8 a_{12}\rho a_1^2 \e^{2\theta_1}+ 4 \e^{\theta_1}((a_1^2- b_1^2) r_1 - 2 a_1 b_1 r_2)}{ 1 + 2 r_1\e^{\theta_1} + a_{12} \rho \e^{2\theta_1}}-\frac{ 8 \left( a_{12} \rho a_1 \e^{2\theta_1}+\e^{\theta_1} (a_1r_1-b_1r_2)\right)^2}{ \left(1 + 2r_1 \e^{\theta_1}+ a_{12} \rho \e^{2\theta_1} \right)^2},
\end{split}
\end{flalign}
and
\begin{flalign}
\begin{split}
&\rho=\alpha_1^2+\beta_1^2,\,r_1=\alpha_1\cos\eta_1-\beta_1\sin\eta_1,\,r_2=\alpha_1\sin\eta_1+\beta_1\cos\eta_1,\\
&r_3=(a_1^2-b_1^2)\cos\eta_1-2a_1b_1\sin\eta_1,\,r_4=(a_1^2-b_1^2)\cos\eta_1+2a_1b_1\sin\eta_1.
\end{split}
\end{flalign}

\begin{rk}
In section \ref{sec3}, the arm \(B_1\) corresponds to the formulas \(u_B\) and \(\wideparen{u_B}\) (or \(\breve{u_B}\)). 
The difference between \(u_B\) and \(\wideparen{u_B}\) (or \(\breve{u_B}\)) is that the former does not include the term \(\alpha_1^2 + \beta_1^2\), whereas the latter does. 
Similarly, this distinction applies to \(S_3\), and the same holds for \(L_j\) and \(\wideparen{L_j}\) (or \(\breve{L_j}\)) in the following paragraphs.
\end{rk}

\begin{table}
  \centering
  \begin{tabular}{cccccc}
    \Xhline{1pt}
    Arm & Amplitude  & Velocity ($(x,\,y)$-direction)  &Trajectory & Component \\
    \hline
    \multirow{3}{*}{$\mathbf{B_1}$}  & \multirow{3}{*}{$\frac{2a_1^2\sqrt{a_{12}}+2b_1^2}{\sqrt{a_{12}}-1}$} & \multirow{3}{*}{$(v_{[x]}^B,\,v_{[y]}^B)$}&  $L_1$ & $u_{B}$\\
     & &  &$\wideparen{L_1}$ & $\wideparen{u_{B}}$ \\
     & &  &$\breve{L}_1$  & $\breve{u}_B$\\
    \hline
    \multirow{2}{*}{$\mathbf{S_3}$}  & \multirow{2}{*}{$\frac{k_3^2}{2}$} & \multirow{2}{*}{$(k_j^2-\frac{3p_j^2}{k_j^2}, \,\frac{k_j^4-3p_j^2}{k_jp_j})$} & $L_3$ & $u_{S}$\\
     & & &$\wideparen{L_3}$ & $\wideparen{u_{S}}$ \\
    \hline
   $\mathbf{S_{1-3}}$  & $\frac{(2a_1-k_3)^2}{2}$ &$(v_{[x]}^{1-3},\,v_{[y]}^{1-3})$ & $L_{1-3}$ & $u_{1-3}$\\
    \hline
   $\mathbf{S_{1+3}}$  & $\frac{(2a_1+k_3)^2}{2}$ &$(v_{[x]}^{1+3},\,v_{[y]}^{1+3})$ & $L_{1+3}$ & $u_{1+3}$\\
    \Xhline{1pt}
  \end{tabular}
  \caption{Physical quantities of the arms in section \ref{sec3} (KPI equation). 
  The relevant formulas are listed by Eqs.\ \eqref{asy11}--\eqref{velocity1} and \eqref{uj04}--\eqref{velocity3}.} \label{tab:t2}
\end{table}

\begin{rk}
Here we still think $u_{1-3}$ Eq. \eqref{asystem1} as a virtual soliton between two  V-shaped breather-soltion. 
This is a very crucial observation for us to determine the trajectory of the stem structure in section \ref{sec3}. 
This virtual soliton has been paid very few  attention comparing with virtual soliton in section \ref{sec2}.
\end{rk}

\begin{figure}[h]
\centering
{\includegraphics[height=4cm,width=4cm]{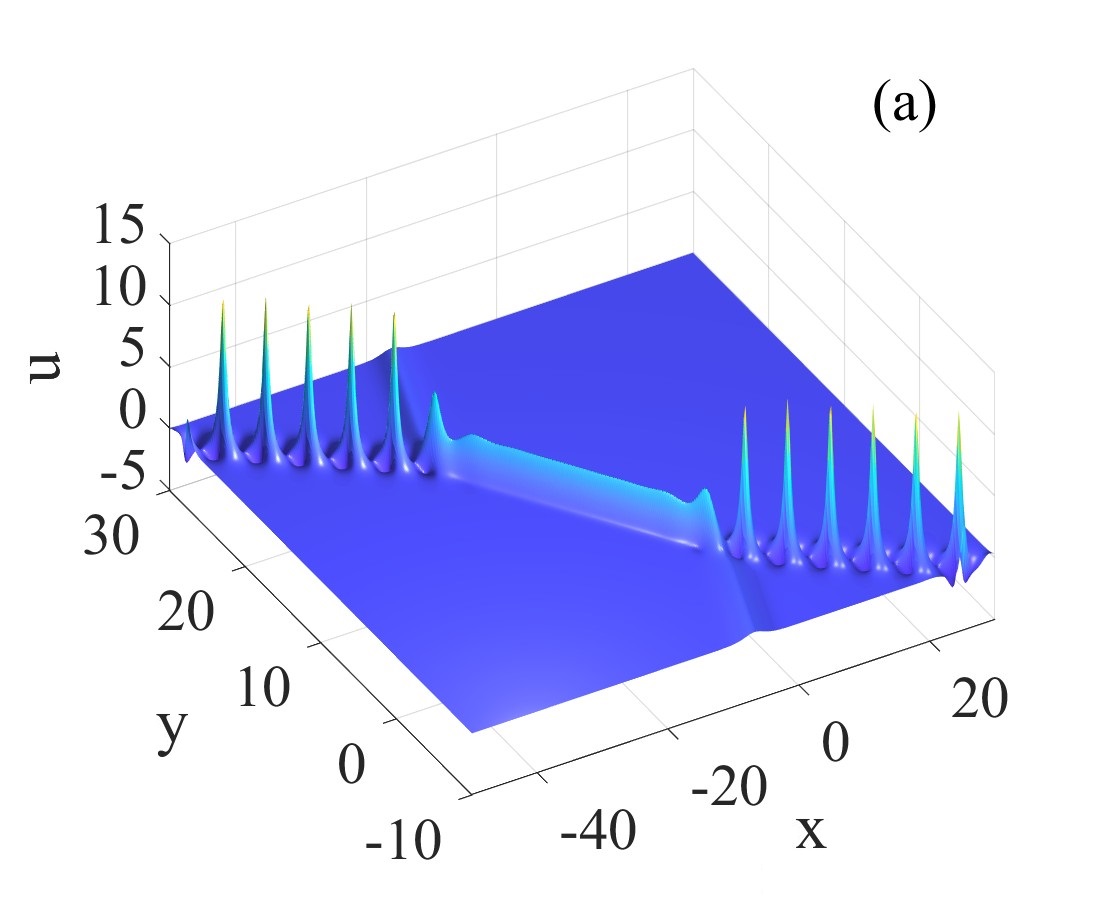}}
{\includegraphics[height=4cm,width=4cm]{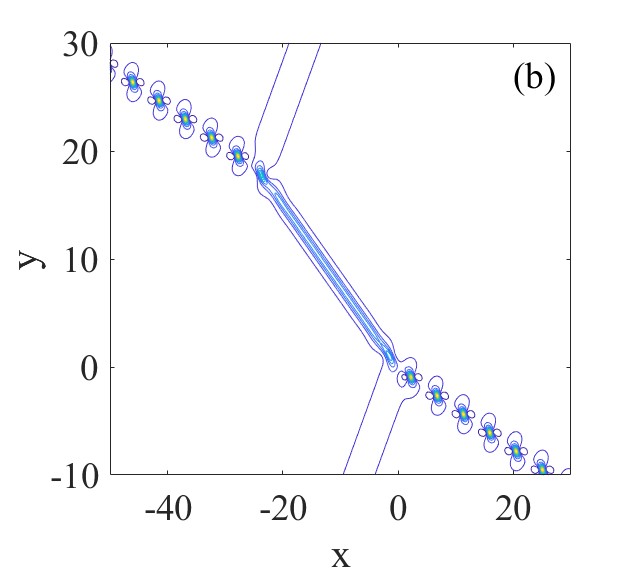}}
{\includegraphics[height=4cm,width=5cm]{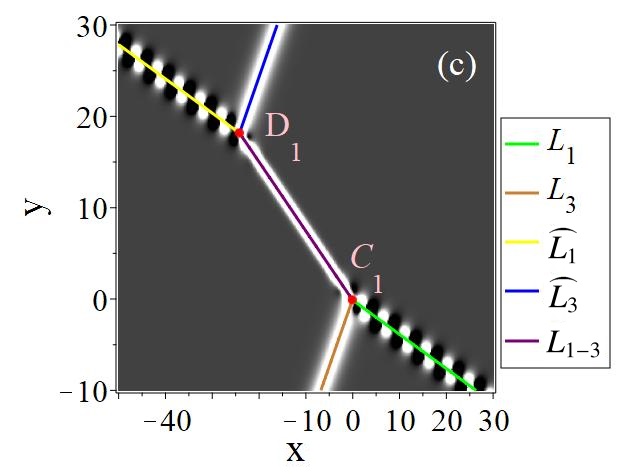}}
{\includegraphics[height=4cm,width=4cm]{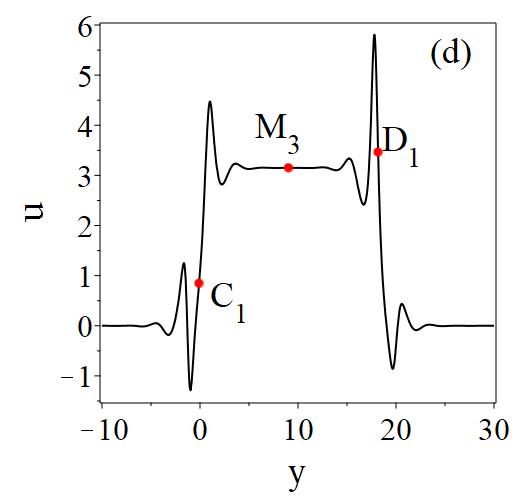}}
\caption{The weakly quasi-resonant breather-soliton \eqref{3u01} with parameters: $k_3 = a_1-\frac{a_1 d_1-b_1 c_1}{a_1^2 + b_1^2}+\epsilon,\,p_3 = a_1b_1+c_1-\frac{(a_1 d_1 - b_1 c_1) (a_1 c_1 + b_1 d_1)}{(a_1^2 + b_1^2)^2},\,a_1=\frac{3}{4},\,b_1=-1,\,c_1=2,\,d_1=1,\,\epsilon=10^{-8},\,t=0$. 
(a) 3D map; (b) Contour plot; (c) The density plot and trajectories; (d) The section-cross curve $u|_{L_{1-3}}$.}\label{fig-bs-1}
\end{figure}

The weakly quasi-resonant breather-soliton $u_{b-s}$ exhibits five arms, similar to the quasi-resonant soliton (see Fig.\ \ref{fig-bs-1}). 
Specifically, it features four infinitely extended arms arranged in two pairs of V-shaped structures, with a local structure connecting these arms at the center. 
The asymptotic forms for the two pairs of V-shaped structures are provided by Eqs.\ \eqref{xasy11}--\eqref{xasy14}. 
The trajectories, amplitudes, and velocities in the $(x,\,y)$-direction for each arm are summarized in Table \ref{tab:t2}, where
\begin{flalign}\label{l03}
\begin{split}
&\boldsymbol{L_1:}\quad \theta_1+\frac{1}{2}\ln a_{12}=0,\quad\,\boldsymbol{\wideparen{L_1}:}\quad \theta_1+\frac{1}{2}\ln(\alpha_1^2+\beta_1^2)+\frac{1}{2}\ln a_{12}=0,\\
&\boldsymbol{L_3:}\quad \xi_3=0,\quad\boldsymbol{\wideparen{L_3}:}\quad \xi_3+\ln(\alpha_1^2+\beta_1^2)=0,\quad\boldsymbol{L_{1-3}:}\quad 2\theta_1-\xi_3+\ln a_{12}=0,
\end{split}
\end{flalign}
and
\begin{flalign}\label{velocity1}
\begin{split}
&v^{B}_{[x]}=a_1^2-3b_1^2-\frac{3(c_1^2-d_1^2)}{a_1^2+b_1^2}-\frac{6b_1c_1d_1}{a_1(a_1^2+b_1^2)},\,v^{B}_{[y]}=\frac{a_1^3-3a_1b_1^2}{c_1}-\frac{3a_1c_1^2-3a_1d_1^2+6b_1c_1d_1}{c_1(a_1^2+b_1^2)},\\
&v^{1-3}_{[x]}=-\frac{k_1^3}{2a_1-k_3}+\frac{3p_1^3}{k_1(2a_1-k_3)}+\frac{2a_1^3-6a_1b_1^2}{2a_1-k_3}-\frac{6a_1(c_1^2-d_1^2)+12b_1c_1d_1}{(a_1^2+b_1^2)(2a_1-k_3)},\\
&v^{1-3}_{[y]}=-\frac{k_1^3}{2c_1-p_3}+\frac{3p_1^3}{k_1(2c_1-p_3)}+\frac{2a_1^3-6a_1b_1^2}{2c_1-p_3}-\frac{6a_1(c_1^2-d_1^2)+12b_1c_1d_1}{(a_1^2+b_1^2)(2c_1-p_3)}.
\end{split}
\end{flalign}

\begin{figure}[h]
\centering
{\includegraphics[height=4cm,width=4cm]{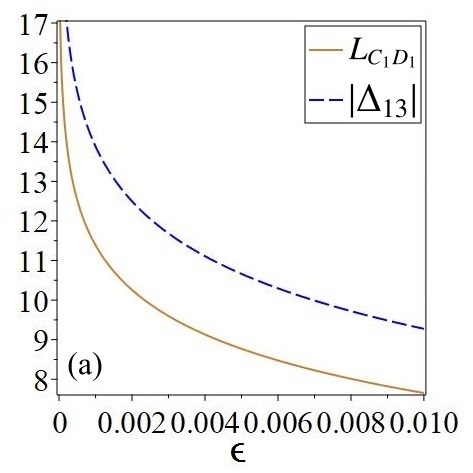}}
{\includegraphics[height=4cm,width=6cm]{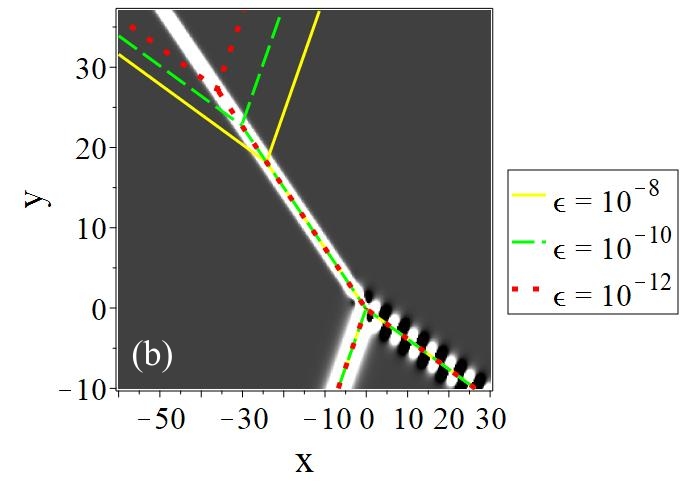}}
{\includegraphics[height=4cm,width=5.7cm]{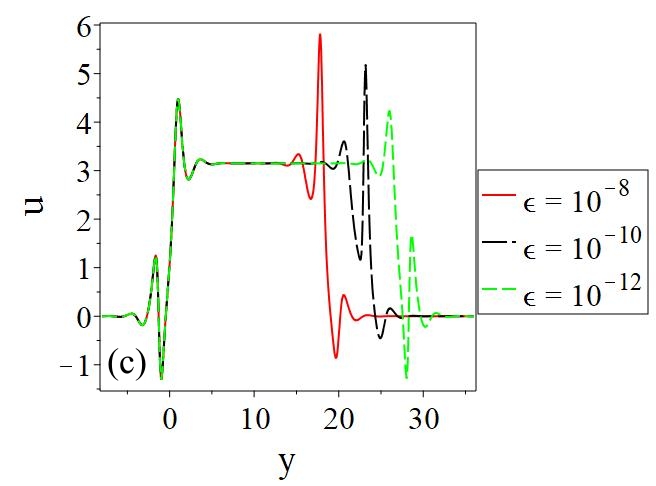}}\\
{\includegraphics[height=4cm,width=8cm]{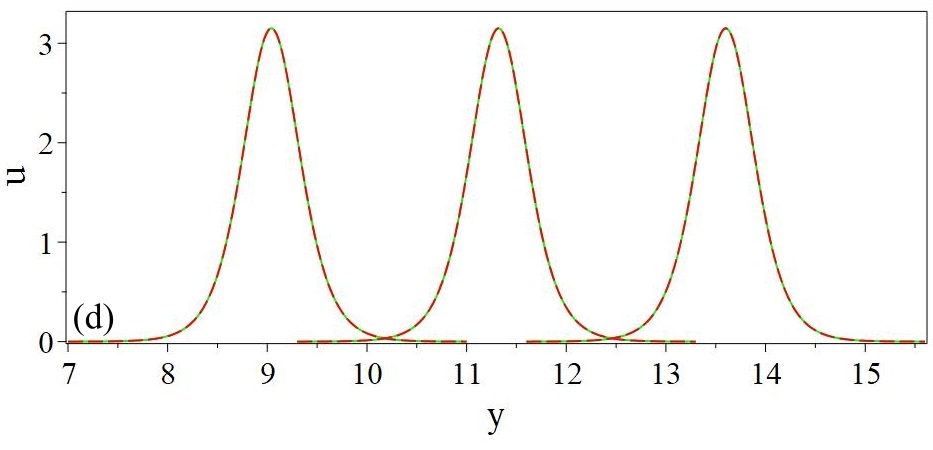}}
\caption{Parameters: $k_3 = a_1-\frac{a_1 d_1-b_1 c_1}{a_1^2 + b_1^2}+\epsilon,\,p_3 = a_1b_1+c_1-\frac{(a_1 d_1 - b_1 c_1) (a_1 c_1 + b_1 d_1)}{(a_1^2 + b_1^2)^2},\,a_1=\frac{3}{4},\,b_1=-1,\,c_1=2,\,d_1=1,\,t=0$. 
(a) Graphs of \( L_{C_1D_1} \) and \(|\Delta_{13}| \) as the function of \( \epsilon \); (b) The trajectories of \eqref{3u01} with different $\epsilon$; (c) The section curves $u|_{L_{1-3}}$ with different $\epsilon$; (d) The section curves that passes through \(M_3\) and is perpendicular to the trajectory \(L_{1-3}\) with different $\epsilon$, and from left to right correspond to \(\epsilon = 10^{-8},\, 10^{-10},\, 10^{-12}\), and the green curves correspond to $u|_{L_{1-3}}$ while the red curves correspond to $u_{1-3}$. Note in (d) that this is a combination picture  of  three sectional curves associated with different values of $\epsilon$,  which does not represent any periodic profile. }\label{fig-bs-2}
\end{figure}

It is evident that the lines \(L_1\), \(L_3\), and \(L_{1-3}\) intersect at one point, while the lines \(\wideparen{L_1}\), \(\wideparen{L_3}\), and \(L_{1-3}\) intersect at another point. 
We define these two intersection points as the endpoints of the stem structure, denoting them as \(C_1\) and \(D_1\), respectively. 
The intersections of \(L_1\) with \(L_3\) and of \(\wideparen{L_1}\) with \(\wideparen{L_3}\), as illustrated in Fig.\ \ref{fig-bs-1} (c), are given by:
\begin{flalign}
&C_1:\, \bigg(\frac{-p_3\ln a_{12}}{2(a_1p_3-c_1k_3)}+v_{[x]}^{C}t,\,\frac{k_3\ln a_{12}}{2(a_1p_3-c_1k_3)}+v_{[y]}^{C}t\bigg),\label{eqc1}\\
&D_1:\, \bigg(\frac{(2c_1-p_3)\ln\rho-p_3\ln a_{12}}{2(a_1p_3-c_1k_3)}+v_{[x]}^{C}t,\,\frac{k_3\ln a_{12}-(2a_1-k_3)\ln\rho}{2(a_1p_3-c_1k_3)}+v_{[y]}^{C}t\bigg),\label{eqd1}
\end{flalign}
where,
\begin{flalign}
\begin{split}\label{velocity2}
&v_{[x]}^{C}=\frac{c_1(2p_3^2-k_3^4)}{2(a_1p_3-c_1k_3)}+\frac{a_1p_3(a_1^2-3b_1^2)}
{k_3(a_1p_3-c_1k_3)}-\frac{3p_3(a_1c_1^2-a_1d_1^2+2b_1c_1d_1)}{(a_1^2+b_1^2)(a_1p_3-c_1k_3)},\\
&v_{[y]}^{C}=\frac{a_1k_3(k_3^2-a_1^2+3b_1^2)}{a_1p_3-c_1k_3}-\frac{3a_1p_3^2}
{k_3(a_1p_3-c_1k_3)}+\frac{3k_3(a_1c_1^2+a_1d_1^2-2b_1c_1d_1)}{(a_1^2+b_1^2)(a_1p_3-c_1k_3)}.
\end{split}
\end{flalign}
Consequently, the length of the stem, denoted as $L_{C_1D_1}$, is expressed by:
\begin{equation}\label{eqcd1}
L_{C_1D_1} = \left| \frac{\ln \rho}{2(a_1p_3-c_1k_3)} \right| \sqrt{(2a_1-k_3)^2+(2c_1-p_3)^2}.
\end{equation}
Fig.\ \ref{fig-bs-2} (a) illustrates the relationship between \( L_{C_1D_1} \) and the phase shift \(|\Delta_{13}|\) as functions of \(\epsilon\), where $\epsilon$ comes from the specific form of $k_3$ in order to implement weakly quasi-resonant condition. 
Both the formulas and the figures demonstrate that as \(\epsilon\) decreases, \(L_{C_1D_1}\) and \(|\Delta_{13}|\) increase. 
In the limit as \(\epsilon \to 0\), both \(|\Delta_{13}|\) and \(L_{C_1D_1}\) approach infinity, resulting in the transformation of the breather-soliton into a weakly resonant breather-soliton (Y-shaped breather-soliton). 
Thus, the weakly resonant breather-soliton represents the limiting case of the weakly quasi-resonant breather-soliton, which is itself an intermediate state between the X-shaped and Y-shaped breather-solitons.

The trajectories of weakly quasi-resonant breather-solitons for various values of \(\epsilon\) are presented in Fig.\ \ref{fig-bs-2} (b). 
The background plane shows a density map of the weakly resonant breather-soliton solution \eqref{3weakresonance}. 
As depicted, each configuration features a pair of V-shaped breather-solitons connected by a central stem structure extending in both the northeast and southwest directions. 
As \(\epsilon\) decreases, the V-shaped breather-soliton in the northeast moves further from the V-shaped breather-soliton in the southwest, thereby lengthening the stem structure. 
When \(\epsilon = 0\), the stem structure becomes infinitely long, and the quasi-resonant breather-soliton transitions into the resonant breather-soliton represented by the background plane.

Next, we analyze the cross-section of the stem \(S_{1-3}\). 
The cross-sectional curve of the weakly quasi-resonant breather-soliton \eqref{3u01} on the plane defined by $L_{1-3}$: \(2\theta_1 - \xi_3 + \ln a_{12} = 0\) is examined as follows:
\begin{equation}
\begin{aligned}\label{crossbs01}
&u|_{L_{1-3}}=
\frac{4g_5a_{12}\e^{-\Theta_3}+2(2a_1+k_3)\rho a_{12}^2+2(4a_1^2+k_3^2)a_{12}\e^{-2\Theta_3}+4g_4\e^{-3\Theta_3}}{\e^{-4\Theta_3} + 2 \e^{-3\Theta_3} \cos\Theta_4 + 2 a_{12} \e^{-2 \Theta_3} + 2 a_{12} \e^{-\Theta_3} g_1 + a_{12}^2\rho} \\
&\quad- \frac{2\Big(2a_{12}((a_1+k_3)g_1-b_1g_2)\e^{-\Theta_3}+(2a_1+k_3)\rho a_{12}^2+(2a_1+k_3)a_{12}\e^{-2\Theta_3}+2g_3 \e^{-3\Theta_3} \Big)^2}{(\e^{-4\Theta_3} + 2 \e^{-3\Theta_3} \cos\Theta_4 + 2 a_{12} \e^{-2 \Theta_3} + 2 a_{12} \e^{-\Theta_3} g_1 + a_{12}^2\rho)^2},
\end{aligned}
\end{equation}
where
\begin{flalign*}
&g_1=\alpha_1 \cos\Theta_4 - \beta_1 \sin\Theta_4,\,g_2=\alpha_1 \sin\Theta_4+\beta_1 \cos\Theta_4,\,g_3=a_1\cos\Theta_4-b_1\sin\Theta_4,\\
&g_4=(a_1^2-b_1^2)\cos\Theta_4-2a_1b_1\sin\Theta_4,\,g_5=((a_1+k_3)^2-b_1^2)g_1-2b_1(a_1+k_3)g_2,\\
&\Theta_3=\frac{a_1p_3-c_1k_3}{2a_1-k_3}y+\bigg(\frac{2a_1^3-6a_1^2b_1^2-a_1k_3^3}{2a_1-k_3}-\frac{6a_1^2(c_1^2-d_1^2)+12a_1b_1c_1d_1}
{(2a_1-k_3)(a_1^2+b_1^2)} \\
&+\frac{3a_1p_3^2}{k_3(2a_1-k_3)}+\frac{3a_1(c_1^2-d_1^2)+6b_1c_1d_1}{a_1^2+b_1^2}-a_1(a_1^2-3b_1^2) \bigg)t-\frac{a_1\ln a_{12}}{2a_1-k_3},\\
&\Theta_4=\bigg(d_1-\frac{b_1(2c_1-p_3)}{2a_1-k_3}\bigg)y+\bigg(\frac{2a_1^2b_1-6a_1b_1^3-b_1k_3^3}{2a_1-k_3}-\frac{6a_1b_1(c_1^2-d_1^2)+12b_1^2c_1d_1}
{(2a_1-k_3)(a_1^2+b_1^2)} \\
&+\frac{3b_1p_3^2}{k_3(2a_1-k_3)}-\frac{3b_1(c_1^2-d_1^2)-6a_1c_1d_1}{a_1^2+b_1^2}-b_1(3a_1^2-b_1^2) \bigg)t-\frac{b_1\ln a_{12}}{2a_1-k_3}.
\end{flalign*}

The cross-sectional curves \( u|_{L_{1-3}} \) for various values of \(\epsilon\) are depicted in Fig.\ \ref{fig-bs-2} (c) with different values of $\epsilon$. 
These curves further confirm that as \(\epsilon\) decreases, the length of the stem increases. 
Calculating the extreme values by differentiating Eq.\ \eqref{crossbs01} proves to be very tedious, making it challenging to precisely determine the exact extreme point of the stem \(S_{1-3}\) between \(C_1\) and \(D_1\). 
As a result, we approximate the extreme point by taking the midpoint of the line segment \(C_1D_1\), denoted as \(M_3\), as illustrated in Fig.\ \ref{fig-bs-1} (c). The coordinates of the point \(M_3\) in the \((x, y)\)-plane are given by:
\begin{flalign}\label{extremr3}
\begin{split}
M_3: \left(\frac{(2c_1 - p_3)\ln \rho - 2p_3 \ln a_{12}}{4(a_1 p_3 - c_1 k_3)} + v_{[x]}^{C}t, \frac{2k_3 \ln a_{12} - (2a_1 - k_3)\ln \rho}{4(a_1 p_3 - c_1 k_3)} + v_{[y]}^{C}t\right).
\end{split}
\end{flalign}

Substituting $M_3$ into equation \eqref{crossbs01}, we get a cumbersome expression of $u(M_3)$, but we are failed to get a simple form of $\underset{\epsilon \to 0}  \lim\,u(M_3)$ as we have done in for the KPII. 
So, we have to compare $u|_{L_{1-3}}$ given by  Eq. \eqref{crossbs01} with $u_{1-3}$  expressed by Eq. \eqref{asystem1} in a numerical way. 
Next, we analyze the cross-sectional curve of the stem structure that passes through \(M_3\) and is perpendicular to the trajectory \(L_{1-3}\), specifically, the cross-sectional curve on the plane defined by \((2c_1 - p_3)(x - x_{M_3}) - (2a_1 - k_3)(y - y_{M_3}) = 0\). 
Fig.\ \ref{fig-bs-2} (d) displays the cross-sectional curves of the stem structure  $u|_{L_{1-3}}$ and the virtual soliton $u_{1-3}$, for various values of \(\epsilon\) which  originates from $k_3$. 
As observed in Fig. \ref{fig-bs-2} (d), the cross-sectional curves of $u|_{L_{1-3}}$ are nearly identical to  profiles of $u_{1-3}$. 
This excellent agreement shows the virtual soliton $u_{1-3}$ is a good approximation of  main part (i.e., almost flap top)  of the stem structure.

\subsection{Stem structure in the strongly quasi-resonant breather-soliton}\label{sec3.3}
In cases where \(\Delta_{13} \approx +\infty\) (equivalently, \(a_{13} \approx +\infty\)), the breather-soliton undergoes strongly quasi-resonant collisions. 
By applying the transformation \(\theta_1 \to \theta_1 - \frac{1}{2} \ln (\alpha_1^2 + \beta_1^2)\) in Eq.\ \eqref{3fbs}, which is equivalent to the transformations \(\xi_1 \to \xi_1 - \ln a_{13}\) and \(\xi_2 \to \xi_2 - \ln a_{23}\) in Eq.\ \eqref{3f}, the strongly quasi-resonant breather-soliton can be expressed as follows:
\begin{flalign}
&u_{b+s}=2(\ln f_{b+s})_{xx},\label{3u02}\\
&f_{b+s}=1 + \frac{2( \alpha_1 \cos\eta_1+\beta_1 \sin\eta_1)}{\alpha_1^2+ \beta_1^2} \e^{\theta_1}+ \e^{\xi_3}  + 2 \e^{\theta_1+\xi_3} \cos\eta_1+ \frac{a_{12}}{\alpha_1^2 + \beta_1^2} \e^{2\theta_1}  + a_{12}  \e^{2\theta_1 + \xi_3}.\label{3f02}
\end{flalign}
\begin{rk}
The strongly quasi-resonant breather-soliton can also be derived directly from the tau function \eqref{3fbs} by setting \(\alpha_1^2 + \beta_1^2 \approx \infty\). 
The transformation \(\theta_1 \to \theta_1 - \frac{1}{2} \ln (\alpha_1^2 + \beta_1^2)\) is applied to ensure consistency with the strongly resonant breather-soliton solution \eqref{3strongf}.
\end{rk}

To obtain the strongly quasi-resonant breather-soliton solution, we first need to identify the conditions that the parameters must be satisfied in the resonant case. Analyzing Eq.\ \eqref{a123}, we find that the resonant condition \(\Delta_{13} = +\infty\) is equivalent to either \(n_1 = n_2 = 0\) or \(n_3 = n_4 = 0\). 
Consequently, the resonant condition can be expressed as follows:
\begin{flalign}
	\begin{split}
k_3 = -a_1-\frac{ a_1 d_1 - b_1 c_1}{a_1^2 + b_1^2},\,
p_3=a_1b_1-c_1-\frac{(a_1 d_1 - b_1 c_1) (a_1 c_1 + b_1 d_1)}{(a_1^2 + b_1^2)^2};
	\end{split}\label{kp03}
\end{flalign}
or
\begin{flalign}
	\begin{split}
k_3 =a_1+\frac{a_1 d_1 - b_1 c_1}{a_1^2 + b_1^2},\,
p_3=-a_1b_1+c_1+\frac{(a_1 d_1 - b_1 c_1) (a_1 c_1 + b_1 d_1)}{(a_1^2 + b_1^2)^2}.
	\end{split}\label{kp04}
\end{flalign}

Thus, the criteria that the parameters must meet for a strongly quasi-resonant collision are as follows:

(1) $k_3 = -a_1-\frac{ a_1 d_1 - b_1 c_1}{a_1^2 + b_1^2}+\epsilon$, \quad (2) $p_3 = a_1b_1-c_1-\frac{(a_1 d_1 - b_1 c_1) (a_1 c_1 + b_1 d_1)}{(a_1^2 + b_1^2)^2}+\epsilon$,

(3) $k_3 = a_1+\frac{ a_1 d_1 - b_1 c_1}{a_1^2 + b_1^2}+\epsilon$, \quad (4) $p_3 = -a_1b_1+c_1+\frac{(a_1 d_1 - b_1 c_1) (a_1 c_1 + b_1 d_1)}{(a_1^2 + b_1^2)^2}+\epsilon$.

Without loss of generality, we deals only with the case (1) $k_3 =-a_1-\frac{ a_1 d_1 - b_1 c_1}{a_1^2 + b_1^2}+\epsilon,\,p_3 = a_1b_1-c_1-\frac{(a_1 d_1 - b_1 c_1) (a_1 c_1 + b_1 d_1)}{(a_1^2 + b_1^2)^2}$. 
When $\alpha_1^2+\beta_1^2\approx +\infty$, it has the following asymptotic forms:

Before collision:

The breather $(\theta_1\approx 0,\,\xi_3\to -\infty)$:
\begin{flalign}
f\sim f_{B}^-= 1+ \frac{2( \alpha_1 \cos\eta_1+\beta_1 \sin\eta_1)}{{\alpha_1}^2 + {\beta_1}^2} \e^{\theta_1}+ \frac{a_{12}}{\alpha_1^2 + \beta_1^2} \e^{2\theta_1}, \label{xasy21}
\end{flalign}

The soliton $(\theta_1\to -\infty,\,\xi_3\approx 0)$:
\begin{flalign}
f\sim f_{S}^-= 1+\e^{\xi_3},\label{xasy22}
\end{flalign}

After collision:

The breather $(\theta_1\approx 0,\,\xi_3\to +\infty)$:
\begin{flalign}
f\sim f_{B}^+= 1+2\e^{\theta_1}\cos\eta_1+a_{12}\e^{2\theta_1},\label{xasy23}
\end{flalign}

The soliton $(\theta_1\to +\infty,\,\xi_3\approx 0)$:
\begin{flalign}
f\sim f_{S}^+= 1+(\alpha_1^2 + \beta_1^2)\e^{\xi_3},\label{xasy24}
\end{flalign}

The stem  $S_{1+3}$ $(2\theta_1\approx -\xi_3,\,\theta_1\to+\infty,\,\xi_3\to -\infty)$:
\begin{flalign}
	\begin{split}
f\sim f_{1+3}= 1+a_{12}\e^{2\theta_1+\xi_3}.
	\end{split}\label{xasy25}
\end{flalign}

Sorting out the above analytical results, the asymptotic form of the solution can be obtained as following proposition:

The strongly quasi-resonant breather-soliton has the following asymptotic forms:\\
Before collision $(x\to-\infty)$:
\begin{flalign}
\begin{split}
&\mathbf{B_1}:\,\, f\sim  1+ \frac{2( \alpha_1 \cos\eta_1+\beta_1 \sin\eta_1)}{{\alpha_1}^2 + {\beta_1}^2} \e^{\theta_1}+ \frac{a_{12}}{\alpha_1^2 + \beta_1^2} \e^{2\theta_1},\,u\sim \breve{u}_B,\\
&\mathbf{S_3}:\,\, f\sim 1+\e^{\xi_3},\,u\sim u_{S}=\frac{k_3^2}{2}\sech(\xi_3);
\end{split}\label{asy21}
\end{flalign}
After collision $(x\to+\infty)$:
\begin{flalign}
\begin{split}
&\mathbf{B_1}:\,\,f\sim 1+2\e^{\theta_1}\cos\eta_1+a_{12}\e^{2\theta_1},\, u\sim u_B,\\
&\mathbf{S_3}:\,\, f\sim 1+(\alpha_1^2 + \beta_1^2)\e^{\xi_3},\,u\sim \wideparen{u_{S}}=\frac{k_3^2}{2}\sech(\xi_3+\ln \rho);\\
\end{split}\label{asy22}
\end{flalign}
The stem structure:
\begin{flalign}
\begin{split}
\mathbf{S_{1+3}}:\,\, f\sim 1+a_{12}\e^{2\theta_1+\xi_3},\,u\sim u_{1+3}=\frac{(2a_1+k_3)^2}{2}\sech(2\theta_1+\xi_3+\ln a_{12}).\label{asystem2}
\end{split}
\end{flalign}
The relevant formulas are given by Eqs. \eqref{asy11}--\eqref{asy12} and
\begin{flalign}\label{uj04}
\breve{u}_B=\frac{8 a_{12} a_1^2 \e^{2\theta_1}+ 4 \e^{\theta_1} \left((a_1^2  -b_1^2)r_2 -2 a_1 b_1r_1\right)}{\rho + 2r_2 \e^{\theta_1} + a_{12}\e^{2\theta_1}}-\frac{ 8\left(a_{12} a_1 \e^{2\theta_1}+ \e^{\theta_1} (a_1 r_2-b_1r_1)\right)^2}{ \left(\rho+ 2 r_2\e^{\theta_1} + a_{12} \e^{2\theta_1} \right)^2}.
\end{flalign}
\begin{rk}
Here  $u_{1+3}$ can also regarded as a virtual soliton.
\end{rk}

The trajectories, amplitudes and the velocities on $(x,\,y)$-direction of each arm are given in Table \ref{tab:t2}, where
\begin{flalign}\label{l04}
\begin{split}
\boldsymbol{\breve{L_1}:}\quad \theta_1-\frac{1}{2}\ln\rho+\frac{1}{2}\ln a_{12}=0,\quad\boldsymbol{L_{1+3}:}\quad 2\theta_1+\xi_3+\ln a_{12}=0,
\end{split}
\end{flalign}
and
\begin{flalign}\label{velocity3}
\begin{split}
v^{1+3}_{[x]}=\frac{k_1^3}{2a_1+k_3}-\frac{3p_1^3}{k_1(2c_1+p_3)}+\frac{2a_1^3-6a_1b_1^2}{2a_1+k_3}-\frac{6a_1(c_1^2-d_1^2)+12b_1c_1d_1}{(a_1^2+b_1^2)(2a_1+k_3)},\\
v^{1+3}_{[y]}=\frac{k_1^3}{2c_1+p_3}-\frac{3p_1^3}{k_1(2c_1+p_3)}+\frac{2a_1^3-6a_1b_1^2}{2c_1+p_3}-\frac{6a_1(c_1^2-d_1^2)+12b_1c_1d_1}{(a_1^2+b_1^2)(2c_1+p_3)}.
\end{split}
\end{flalign}

\begin{figure}[h]
\centering
{\includegraphics[height=4cm,width=4cm]{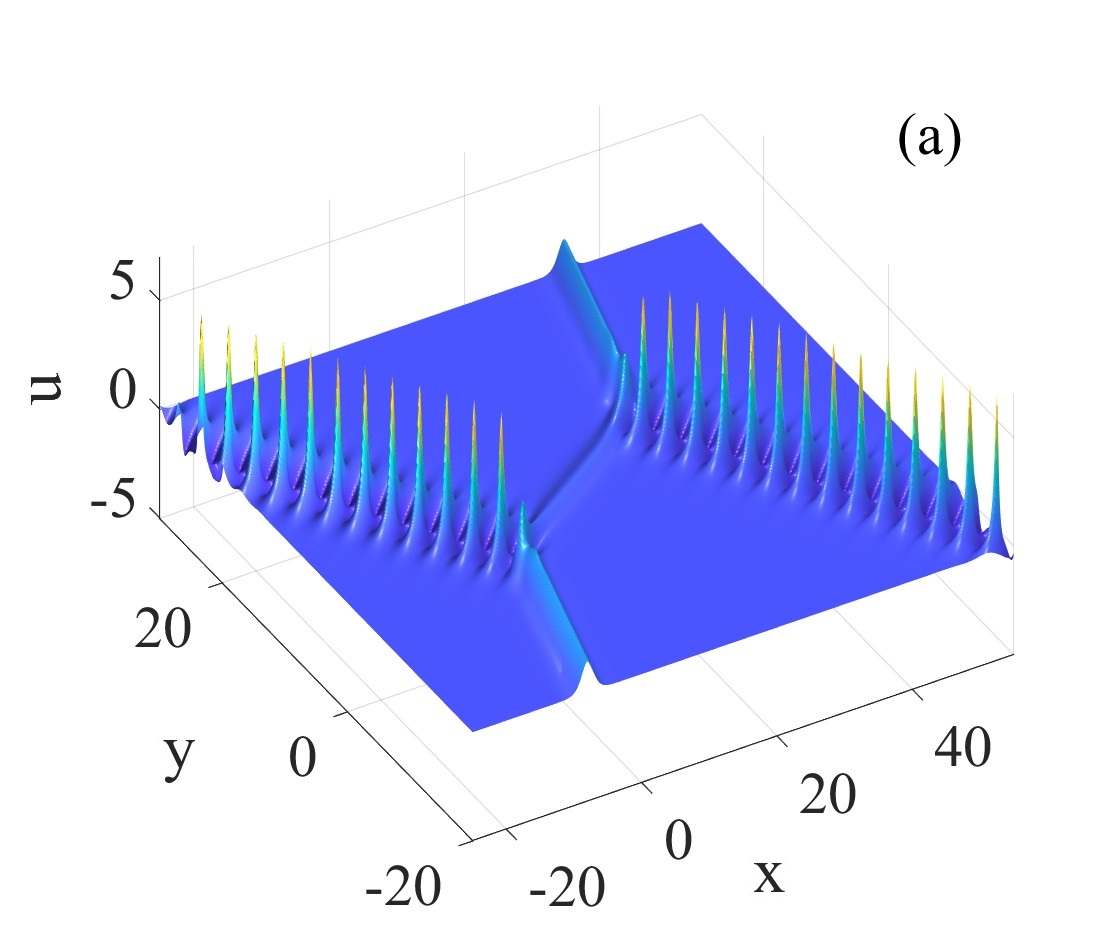}}
{\includegraphics[height=4cm,width=4cm]{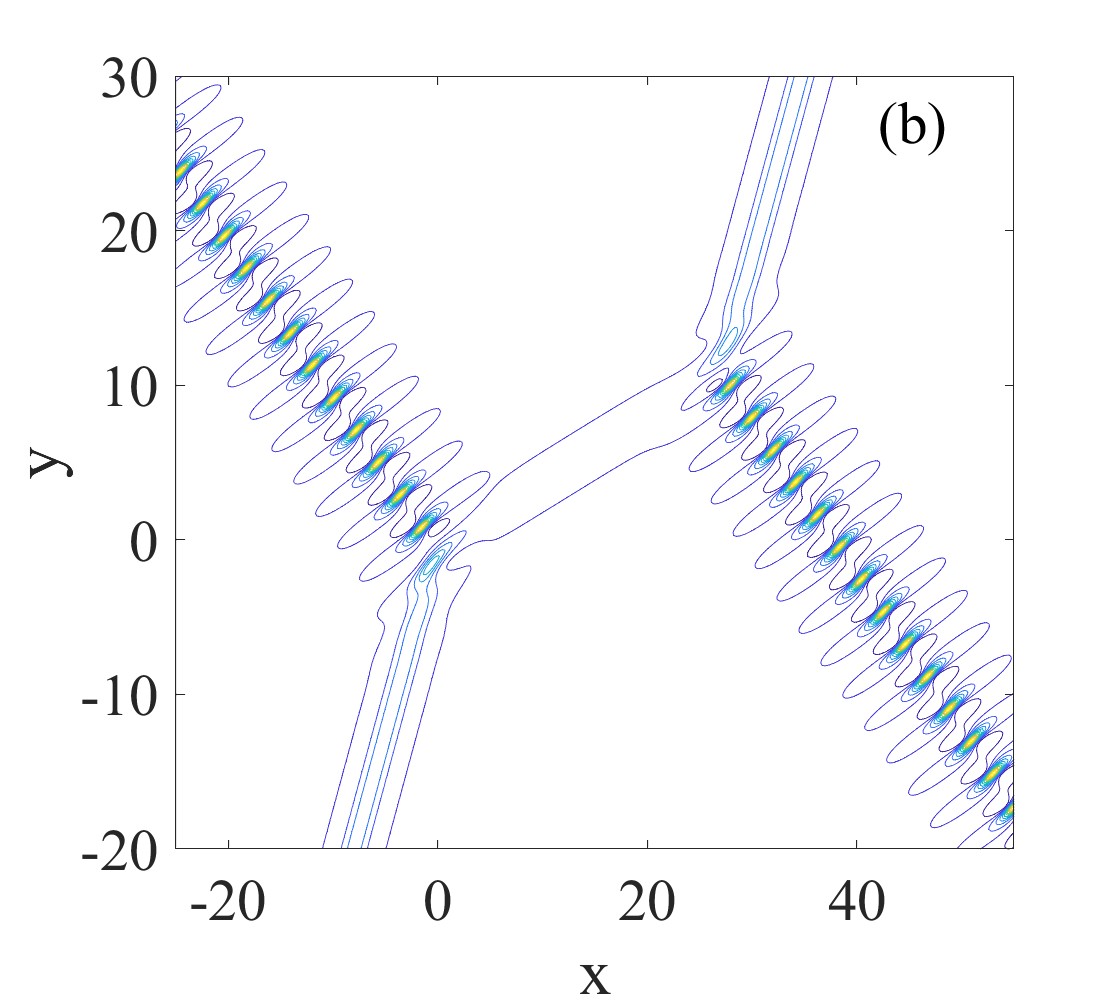}}
{\includegraphics[height=4cm,width=4.7cm]{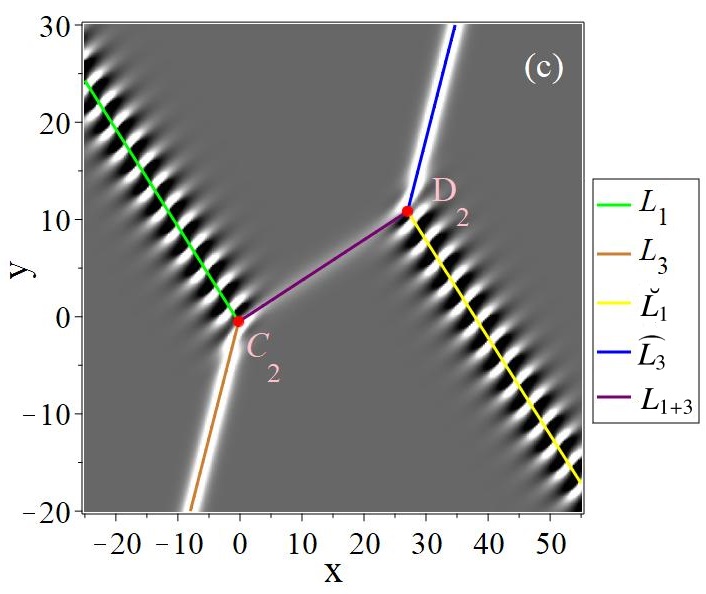}}
{\includegraphics[height=4cm,width=4cm]{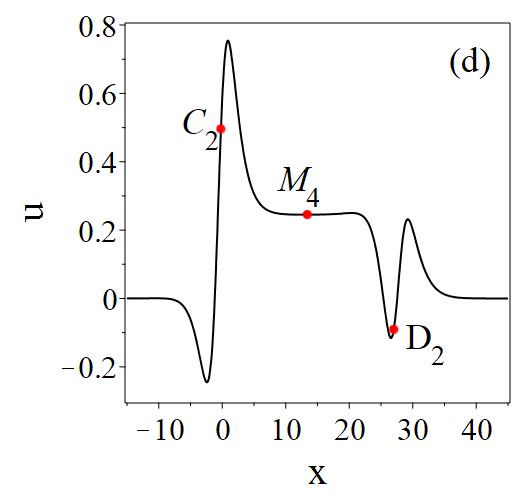}}
\caption{The strongly quasi-resonant breather-soliton \eqref{3u02} with parameters: $k_3 = -a_1-\frac{ a_1 d_1 - b_1 c_1}{a_1^2 + b_1^2}+\epsilon,\,p_3=a_1b_1-c_1-\frac{(a_1 d_1 - b_1 c_1) (a_1 c_1 + b_1 d_1)}{(a_1^2 + b_1^2)^2},\,a_1=\frac{1}{2},\,b_1=-1,\,c_1=\frac{1}{2},\,d_1=2,\,\epsilon=10^{-8},\,t=0$. (a) 3D map; (b) Contour plot; (c) The density plot and trajectories; (d) The section-cross curve $u|_{L_{1+3}}$.}\label{fig-bs-3}
\end{figure}

\begin{figure}[h]
	\centering
  {\includegraphics[height=4cm,width=4cm]{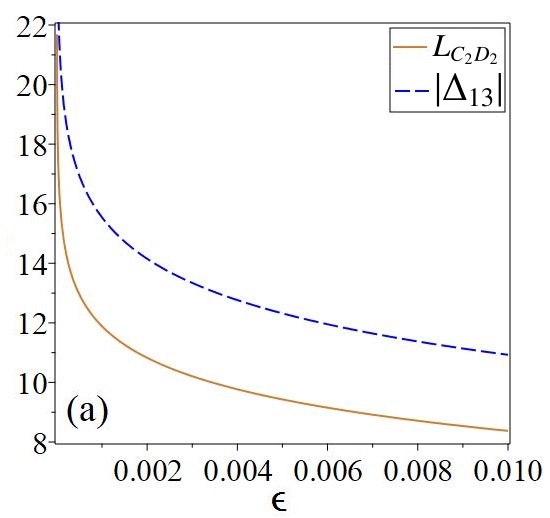}}
  {\includegraphics[height=4cm,width=5.6cm]{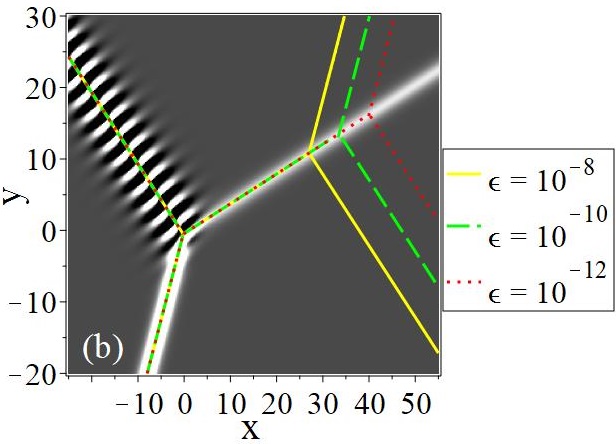}}
  {\includegraphics[height=4cm,width=4cm]{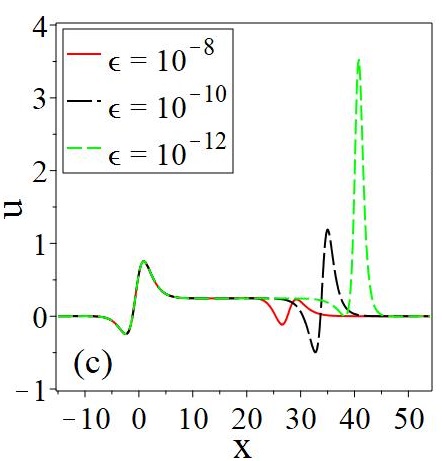}}\\
     {\includegraphics[height=4cm,width=10cm]{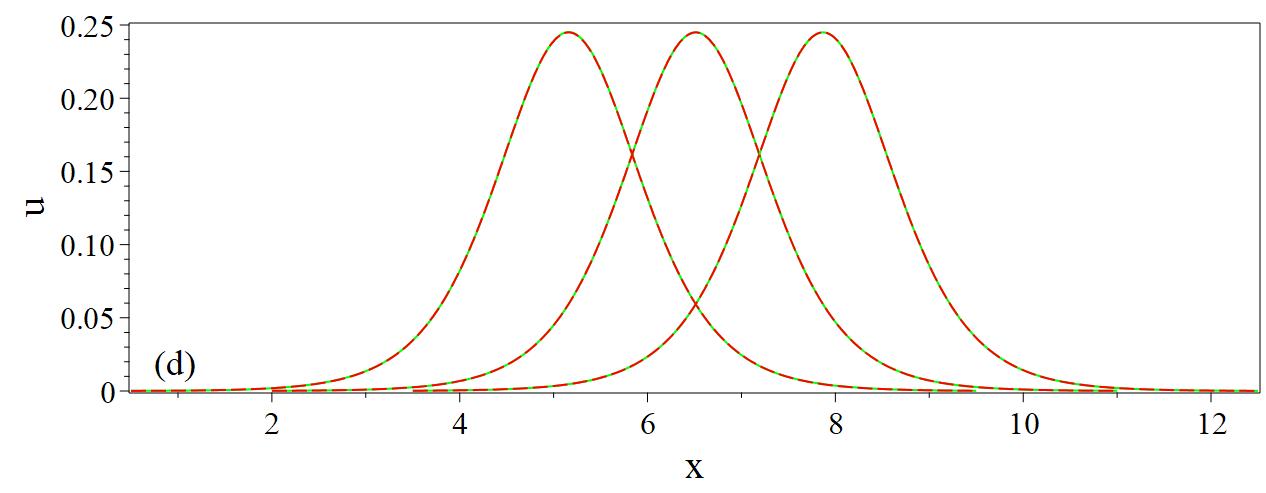}}
	\caption{The strongly quasi-resonant breather-soliton \eqref{3u02} with parameters: $k_3 = -a_1-\frac{ a_1 d_1 - b_1 c_1}{a_1^2 + b_1^2}+\epsilon,\,p_3=a_1b_1-c_1-\frac{(a_1 d_1 - b_1 c_1) (a_1 c_1 + b_1 d_1)}{(a_1^2 + b_1^2)^2},\,a_1=\frac{1}{2},\,b_1=-1,\,c_1=\frac{1}{2},\,d_1=2,\,t=0$. (a) Graphs of \( L_{C_2D_2} \) and \(|\Delta_{13}| \) as the function of \( \epsilon \); (b) The trajectories of \eqref{3u02} with different $\epsilon$; (c) The section curves $u|_{L_{1+3}}$ with different $\epsilon$; (d) The section curves that passes through \(M_4\) and is perpendicular to the trajectory \(L_{1+3}\) with different $\epsilon$, and from left to right correspond to \(\epsilon = 10^{-8},\, 10^{-10},\, 10^{-12}\), and the green curves correspond to $u|_{L_{1+3}}$ while the red curves correspond to $u_{1+3}$. }\label{fig-bs-4}
\end{figure}

It is clear that \(L_1\), \(L_3\), and \(L_{1+3}\) intersect at a single point, whereas \(\breve{L_1}\), \(\wideparen{L_3}\), and \(L_{1+3}\) intersect at a different point. 
These two points of intersection are defined as the endpoints of the stem structure, labeled \(C_2\) and \(D_2\), respectively. 
They are depicted in Fig.\ \ref{fig-bs-3} (c) and have the following form
\begin{flalign}
&C_2:\, \bigg(\frac{-p_3\ln a_{12}}{2(a_1p_3-c_1k_3)}+v_{[x]}^{C}t,\,\frac{k_3\ln a_{12}}{2(a_1p_3-c_1k_3)}+v_{[y]}^{C}t\bigg),\label{eqc1}\\
&D_2:\, \bigg(\frac{(2c_1+p_3)\ln\rho-p_3\ln a_{12}}{2(a_1p_3-c_1k_3)}+v_{[x]}^{C}t,\,\frac{k_3\ln a_{12}-(2a_1+k_3)\ln\rho}{2(a_1p_3-c_1k_3)}+v_{[y]}^{C}t\bigg),\label{eqd1}
\end{flalign}
where $v_{[x]}^{C}$ and $v_{[y]}^{C}$ are given by \eqref{velocity2}. Consequently, the length of the stem, denoted as $L_{C_2D_2}$, is defined as:
\begin{equation}\label{eqcd2}
L_{C_2D_2} = \left| \frac{\ln \rho}{2(a_1p_3-c_1k_3)} \right| \sqrt{(2a_1+k_3)^2+(2c_1+p_3)^2}.
\end{equation}
Fig.\ \ref{fig-bs-4} (a) illustrates how \( L_{C_2D_2} \) and the phase shift \(|\Delta_{13}|\) vary with \(\epsilon\) which originates from $k_3$. 
Both the equations and graphical representations indicate that as \(\epsilon\) decreases, \(L_{C_2D_2}\) and \(|\Delta_{13}|\) increase. 
In the limit where \(\epsilon \to 0\), \(\rho\) approaches zero, causing \(L_{C_2D_2}\) to become infinitely large and resulting in the transformation of the breather-soliton into a strongly resonant breather-soliton (Y-shaped breather-soliton). 
Hence, the strongly resonant breather-soliton is the extreme case of the strongly quasi-resonant breather-soliton, which itself represents an intermediate form between the X-shaped and Y-shaped breather-solitons.

Fig.\ \ref{fig-bs-4} (b) shows the trajectories of strongly quasi-resonant breather-solitons for various \(\epsilon\) values, with the background illustrating a density map of the strongly resonant breather-soliton solution \eqref{3strongresonance}. 
The figure features pairs of V-shaped breather-solitons connected by a central stem structure extending in the northeast and southwest directions. 
As \(\epsilon\) decreases, the V-shaped breather-solitons move further apart, elongating the stem structure. 
When \(\epsilon = 0\), the stem structure becomes infinitely long, transitioning the quasi-resonant breather-soliton to the resonant breather-soliton shown in the background.

Next, we analyze the cross-sectional curve of the stem \(S_{1+3}\). 
This involves examining the cross-sectional curve of the breather-soliton \eqref{3u02} on the plane defined by \(2\theta_1 + \xi_3 + \ln a_{12} = 0\).
\begin{equation}
\begin{aligned}\label{crossbs02}
&u|_{L_{1+3}}=
\frac{2k_3^2 \rho+4h_7\rho\e^{\Theta_5}+2(2a_1+k_3)^2\rho a_{12}\e^{2\Theta_5}+4h_4a_{12}\e^{3\Theta_5}+8a_1^2a_{12}^2\e^{4\Theta_5}}{\rho + 2\rho\cos\Theta_6 \e^{\Theta_5}+ 2 a_{12} \rho \e^{2 \Theta_5} + 2 h_1 \e^{3 \Theta_5} + a_{12}^2\e^{4 \Theta_5}} \\
&\quad- \frac{2\bigg(k_3\rho +h_6a_{12}\rho\e^{\Theta_5}+(2a_1+k_3)\rho a_{12}\e^{2\Theta_5}+2a_{12}h_5\e^{3\Theta_5} +2a_1a_{12}^2\e^{4\Theta_5}\bigg)^2}{(\rho + 2 \rho\cos\Theta_6 \e^{\Theta_5}+ 2 a_{12} \rho \e^{2 \Theta_5} + 2 h_1 \e^{3 \Theta_5} + a_{12}^2\e^{4 \Theta_5})^2}
\end{aligned}
\end{equation}
where
\begin{flalign*}
&h_1=\alpha_1 \cos\Theta_4 + \beta_1 \sin\Theta_4,\,h_2=\alpha_1 \sin\Theta_4+\beta_1 \cos\Theta_4,\,h_3=\alpha_1 \sin\Theta_4-\beta_1 \cos\Theta_4,\\
&h_4=(a_1^2-b_1^2)h_1-2a_1b_1h_2,\,h_5=a_1h_1-b_1h_3,\,h_6=(a_1+k_3)\cos\Theta_4-b_1\sin\Theta_4,\\
&h_7=((a_1+k_3)^2-b_1^2)\cos\Theta_4-2b_1(a_1+k_3)\sin\Theta_4,\\
&\Theta_5=\frac{a_1p_3-c_1k_3}{2c_1+p_3}x+\bigg(\frac{a_1^3-3a_1b_1^2p_3+c_1k_3^3}{2c_1+p_3}+\frac{3c_1p_3^2}{k_3(2c_1+p_3)}-\frac{3a_1p_3(c_1^2+d_1^2)-6b_1c_1d_1p_3}
{(2c_1+p_3)(a_1^2+b_1^2)} \bigg)t-\frac{c_1\ln a_{12}}{2c_1+p_3},\\
&\Theta_6=\bigg(b_1-\frac{d_1(2a_1+k_3)}{2c_1+p_3}\bigg)x+\bigg(\frac{2a_1^3d_1-6a_1b_1^2d_1+d_1k_3^3}{2c_1+p_3}-\frac{6a_1d_1(c_1^2-d_1^2)+12b_1c_1d_1^2}
{(2c_1+p_3)(a_1^2+b_1^2)} \\
&-\frac{3d_1p_3^2}{k_3(2c_1+p_3)}-\frac{3b_1(c_1^2-d_1^2)-6a_1c_1d_1}{a_1^2+b_1^2}-b_1(3a_1^2-b_1^2) \bigg)t-\frac{d_1\ln a_{12}}{2c_1+p_3}.
\end{flalign*}
Figure \ref{fig-bs-4} (c) displays the cross-sectional curves \(u|_{L_{1+3}}\) for different values of \(\epsilon\), demonstrating that a smaller \(\epsilon\) leads to a longer stem. 
Determining the precise extreme point of the stem \(S_{1+3}\) between \(C_2\) and \(D_2\) is challenging due to the complexity of calculating the extreme values from the derivative of Eq.\ \eqref{crossbs02}. 
As an approximation, the midpoint of the line segment \(C_2D_2\), denoted \(M_4\), is used. 
This midpoint is illustrated in Fig.\ \ref{fig-bs-3} (c), and its coordinates in the \((x, y)\)-plane are specified as follows:
\begin{flalign}\label{extremr4}
\begin{split}
M_4:\,\bigg(\frac{(2c_1+p_3)\ln\rho-p_3\ln a_{12}}{4(a_1p_3-c_1k_3)}+v_{[x]}^{C}t,\,\frac{k_3\ln a_{12}-(2a_1+k_3)\ln\rho}{4(a_1p_3-c_1k_3)}+v_{[y]}^{C}t\bigg).
\end{split}
\end{flalign}

We now analyze the cross-sectional curve of the stem structure $u|_{L_{1+3}}$ that intersects at point \(M_4\) and is perpendicular to the trajectory \(L_{1+3}\), as we have done in section \ref{sec3.2}. 
This cross-sectional curve is situated on the plane defined by \((2c_1 + p_3)(x - x_{M_4}) - (2a_1 + k_3)(y - y_{M_4}) = 0\). 
Figure \ref{fig-bs-4} (d) shows the cross-sectional curves of the stem structure $u|_{L_{1+3}}$ along with the virtual soliton $u_{1+3}$, given by \eqref{asystem2}, for different values of \(\epsilon\). 
Here $\epsilon$ originates from $k_3$. 
The figure reveals that these cross-sectional curves of $u|_{L_{1+3}}$ are nearly identical to profiles of $u_{1+3}$. 
This excellent agreement shows the virtual soliton $u_{1+3}$ Eq. \eqref{asystem2} is a good approximation of  main part (i.e., almost flap top)  of the stem structure Eq. \eqref{crossbs02}.

\section{Conclusions}\label{summary}
In this study, we analyzed quasi-resonant collisions, a class of elastic interactions characterized by a finite yet effectively very large phase shift, thus occupying an intermediate regime between standard elastic collisions (finite phase shifts) and fully resonant collisions (infinite phase shifts).  
Our investigation centered on the quasi-resonant solutions of the KPI and KPII equations, with particular emphasis on the detailed structure of the stem---a wave pattern whose quantitative features had not been systematically explored in previous research.

For the KPII equation, we have shown that quasi-resonant two-soliton interactions generate localized stem structures and have provided a comprehensive asymptotic description of their graphic and dynamical properties.
In particular, explicit analytical expressions have been derived for the trajectories, endpoint coordinates, amplitudes, velocities, lengths, and profile curves of stems. 
The corresponding formulas in the weakly and strongly quasi-resonant regimes are given in Eqs.~\eqref{cross2s01} and \eqref{cross2s02}, respectively.
We further demonstrated that the virtual solitons ($u_{1-2}$ and $u_{1+2}$) give accurate approximations to the dominant components of the stem (bottom or top flap) in both regimes.

By applying the quasi-resonant two-soliton model to nearshore water-wave formation, we reproduced the V-shaped patterns connected by a stem observed along coastlines, as discussed in Section~\ref{sec2.3}. 
The explicit analytical formulas allow us to determine precisely the locations of the branch and stem structures, thereby providing a quantitative theoretical framework for interpreting the oceanic observations reported in \cite{kppre2012,photo,jpa2014}. 
Other wave patterns observed in \cite{kppre2012,photo,jpa2014} can be explained in terms of resonant three-soliton or higher-order solutions, which will be investigated in future work.

In addition, in comparison with Refs. \cite{jpa2004,Dimakis2011}, the main novelty of the present work is the first establishment of a detailed, quantitative, and unified analytical framework for the stem structures of quasi-resonant solutions to the KPII equation.
In summary, Ref. \cite{jpa2004} provides an efficient classification of solution patterns for the KPII equation and offers a qualitative illustration of stem locations through graphical representations. 
Although the approach is, in principle, capable of determining the positions of the stems, no systematic analysis or explicit analytical expressions for the stem structures were derived in Ref. \cite{jpa2004}.
On the other hand, Ref. \cite{Dimakis2011} classifies solitons according to the number of phase points; in particular, Figs. 22 and 23 correspond to the case $M=3$, i.e., four phases.
In our work, the soliton order is defined by the integer $N$ in Eq. \eqref{2f}.
The quasi-resonant line solitons of the KPII equation considered here involve three phases, which correspond to the case $M=2$ in Ref. \cite{Dimakis2011}.
Although Ref. \cite{Dimakis2011} derived a general expression for soliton trajectories for $M>1$, it did not provide a detailed correspondence for the arm---stem configuration in the specific quasi-resonant case studied here, nor did it present analytical results for the stem in terms of their length, amplitude, and profile.
These detailed investigations constitute the core contribution of the present paper.

For the KPI equation, where quasi-resonant two-soliton solutions do not exist, we instead examined quasi-resonant breather---soliton interactions.  
We constructed the quasi-resonant breather--soliton solution in both weakly and strongly quasi-resonant cases, given in Eqs.~\eqref{3u01} and \eqref{3u02}, and analyzed their corresponding stem structures using an approach parallel to the KPII case.  
Analytical descriptions of the stem profile are provided in Eqs.~\eqref{crossbs01} and \eqref{crossbs02}.  
We showed that the virtual solitons ($u_{1-3}$ and $u_{1+3}$) offer effective approximations of the main stem components in these cases.  
Moreover, our results confirm that the quasi-resonant breather is, like the quasi-resonant soliton, a traveling wave whose stem length remains constant over time, as demonstrated in Eqs.~\eqref{eqcd1} and \eqref{eqcd2}.  
A notable distinction from the KPII two-soliton case is that the quasi-resonant breather---soliton lacks symmetry about the midpoint \(M_3\) (or \(M_4\)) of the stem \(S_{1-3}\) (or \(S_{1+3}\)).

Finally, we verified that the resonant Y-shaped solution arises as the limiting configuration of the general X-shaped solution as the parameter $\epsilon \to 0$.  
This transition is illustrated in subfigure (b) of Figures~\ref{fig2s-2}, \ref{fig2s-4}, \ref{fig-bs-2}, and \ref{fig-bs-4}.  

Overall, the principal novelty of this work lies in providing the first detailed, quantitative, and unified
analytical framework for stem structures in quasi-resonant KPI and KPII solutions. 
While previous studies identified the existence of quasi-resonant interactions, the precise graphic and dynamical features of
the stem ---including its length, amplitude, trajectory, and profile curves---had not been systematically analyzed.
The results presented here fill this gap and establish a foundation for further analytical interpretation
of quasi-resonant wave patterns observed both theoretically and practically.

\section*{Appendix A}
\addcontentsline{toc}{section}{Appendix A}
To obtain the weakly resonant 2-soliton, we need to ensure $a_{12}=0$, which is equivalent to \( p_2 = \frac{k_2(k_1^2 - k_1k_2 + p_1)}{k_1} \) or \( p_2 = -\frac{k_2(k_1^2 - k_1k_2 - p_1)}{k_1} \). 
Then the weakly resonant 2-soliton of the KPII equation is given by
\begin{equation}\label{weakresonance}
f_{weak}^{[2]}=1+\exp\xi_1+\exp\xi_2,\,u_{weak}^{[2]}=2(\ln f_{weak}^{[2]})_{xx}.
\end{equation}

To obtain the strongly resonant 2-soliton, we do the transformation $\xi_1\to\xi_1+\ln a_{12}$ in Eq.\ \eqref{2f} and take the limit $a_{12}=+\infty$, the strongly resonant 2-soliton of the KPII equation is given by
\begin{equation}\label{strongresonance}
f_{strong}^{[2]}=1+\exp\xi_1+\exp(-\xi_2),\,u_{strong}^{[2]}=2(\ln f_{strong}^{[2]})_{xx}.
\end{equation}

\section*{Appendix B}
\addcontentsline{toc}{section}{Appendix B}
Substituting $\alpha_1=\beta_1=0$ in to \eqref{3f01}, the tau function of weakly resonant breather-soliton solution of KPI equation (Eq. \eqref{kpeq} with $\delta=-3$) is
\begin{equation}\label{3weakf}
f_{weak}^{[3]}=1 + 2 \e^{\theta_1} \cos\eta_1 + \e^{\xi_3} + a_{12} \e^{2\theta_1}.
\end{equation}
Then the weakly resonant breather-soliton are given by
\begin{flalign}
\begin{split}\label{3weakresonance}
u_{weak}^{[3]}=&\frac{2 \left( 2\e^{\theta_1}( (a_1^2-b_1^2) \cos\eta_1 - 2 a_1 b_1 \sin\eta_1) + k_3^2 \e^{\xi_3} + 4 a_{12} a_1^2 \e^{2\theta_1} \right)}{1 + 2 \e^{\theta_1} \cos\eta_1 + \e^{\xi_3} + a_{12} \e^{2\theta_1}} \\
&- \frac{2 \left( 2\e^{\theta_1}(a_1 \cos\eta_1 - b_1 \sin\eta_1) + k_3 \e^{\xi_3} + 2 a_{12} a_1 \e^{2\theta_1} \right)^2}{\left( 1 + 2 \e^{\theta_1} \cos\eta_1 + \e^{\xi_3} + a_{12} \e^{2\theta_1} \right)^2}.
\end{split}
\end{flalign}

Substituting $\theta_1\to\theta_1-\frac{1}{2}\ln (\alpha_1^2+\beta_1^2)$ in to \eqref{3f01}, and taking limit $\alpha_1^2+\beta_1^2\to+\infty$  the tau function of strongly resonant breather-soliton solution of KPI equation is
\begin{equation}\label{3strongf}
f_{strong}^{[3]}=1 + 2 \e^{\theta_{1} + \xi_{3}} \cos \eta_{1} + \e^{\xi_{3}} + a_{12} \e^{2 \theta_{1} + \xi_{3}}.
\end{equation}
Then the strong-resonant breather-soliton are given by
\begin{flalign}
\begin{split}\label{3strongresonance}
u_{strong}^{[3]}=&\frac{2 \left( a_{12} (2 a_{1} + k_{3})^2 \e^{2 \theta_{1} + \xi_{3}} + 2 \left( ((a_{1} + k_{3})^2 - b_{1}^2) \cos \eta_{1} - 2 b_{1} (a_{1} + k_{3}) \sin \eta_{1} \right) \e^{\theta_{1} + \xi_{3}} + k_{3}^2 \e^{\xi_{3}} \right)}{1 + 2 \e^{\theta_{1} + \xi_{3}} \cos \eta_{1} + \e^{\xi_{3}} + a_{12} \e^{2 \theta_{1} + \xi_{3}}} \\
&- \frac{2 \left( a_{12} (2 a_{1} + k_{3}) \e^{2 \theta_{1} + \xi_{3}} + \left( (2 a_{1} + 2 k_{3}) \cos \eta_{1} - 2 b_{1} \sin \eta_{1} \right) \e^{\theta_{1} + \xi_{3}} + k_{3} \e^{\xi_{3}} \right)^2}{\left(1 + 2 \e^{\theta_{1} + \xi_{3}} \cos \eta_{1} + \e^{\xi_{3}} + a_{12} \e^{2 \theta_{1} + \xi_{3}}\right)^2}.
\end{split}
\end{flalign}
\section*{Appendix C}
\addcontentsline{toc}{section}{Appendix C}
The profile curve of the arm $u_B$ of the weakly quasi-resonant breather-soliton in the plane perpendicular to the trajectory $L_1$ is
\begin{equation}\label{crossb1}
u\big|_{L_1}=\frac{2\bigg(a_{12}a_1^2+\sqrt{a_{12}}(a_1^2-b_1^2)\cos\zeta_1-b_1^2\bigg)}{(\sqrt{a_{12}}+\cos\zeta_1)^2},
\end{equation}
where $\zeta_1=\frac{a_1d_1-b_1c_1}{a_1}y-\bigg(2b_1(a_1^2+b_1^2)-\frac{6(a_1c_1+b_1d_1)(a_1d_1-b_1c_1)}{a_1(a_1^2+b_1^2)}\bigg)t-\frac{b_1\ln a_{12}}{2a_1}$. 
Then we can obtain the period of $B_1$ ($u_{B}$ and $\wideparen{u_{B}}$) is $T_{[y]}=\frac{2\pi a_1}{a_1d_1 - b_1c_1}$ on $y$-direction, while $T_{[x]}=-\frac{2\pi c_1}{a_1d_1 - b_1c_1}$ on $x$-direction. 
There amplitude can be obtained as $u_B^{max}=\wideparen{u_{B}}^{max}=\frac{2a_1^2\sqrt{a_{12}}+2b_1^2}{\sqrt{a_{12}}-1}$.

The peaks of $u_{B}$ local at the following points on $(x,\,y)$ plane is:
\begin{equation}\label{peakb1}
\left(-\frac{d_1\ln\sqrt{a_{12}} +c_1\pi}{a_1 d_1 - b_1 c_1}+nT_{[x]}+v^{B}_{[x]}t,\,\frac{b_1\ln\sqrt{a_{12}} +a_1\pi}{2(a_1 d_1 - b_1 c_1)}+nT_{[y]}+v^{B}_{[y]}t\right),
\end{equation}
where $(v^{B}_{[x]},\,v^{B}_{[y]})$ are given by Eq.\ \eqref{velocity1}.

The peaks of $\wideparen{u_{B}}$ local at the following points on $(x,\,y)$ plane is:
\begin{flalign}
\begin{split}\label{peakb2}
&\bigg(-\frac{d_1\ln\sqrt{a_{12}(\alpha_1^2+\beta_1^2)} +c_1(\pi+\arccos\frac{\alpha_1}{\sqrt{\alpha_1^2+\beta_1^2}})}{a_1 d_1 - b_1 c_1}+nT_{[x]}+v^{B}_{[x]}t,\\
&\frac{b_1\ln\sqrt{a_{12}(\alpha_1^2+\beta_1^2)} +a_1(\pi+\arccos\frac{\alpha_1}{\sqrt{\alpha_1^2+\beta_1^2}})}{a_1 d_1 - b_1 c_1}+nT_{[y]}+v^{B}_{[y]}t\bigg).
\end{split}
\end{flalign}

The peaks of $\breve{u_{B}}$ local at the following points on $(x,\,y)$ plane is:
\begin{flalign}
\begin{split}\label{peakb3}
&\bigg(-\frac{d_1(\ln\sqrt{a_{12}}-\ln\sqrt{\alpha_1^2+\beta_1^2})+c_1(\pi+\arccos\frac{\alpha_1}{\sqrt{\alpha_1^2+\beta_1^2}})}{a_1 d_1 - b_1 c_1}+nT_{[x]}+v^{B}_{[x]}t,\\
&\frac{b_1(\ln\sqrt{a_{12}}-\ln\sqrt{\alpha_1^2+\beta_1^2}) +a_1(\pi+\arccos\frac{\alpha_1}{\sqrt{\alpha_1^2+\beta_1^2}})}{a_1 d_1 - b_1 c_1}+nT_{[y]}+v^{B}_{[y]}t\bigg).
\end{split}
\end{flalign}

The peaks of the arm $B_1$ in weakly quasi-resonant is given by \eqref{peakb1} and \eqref{peakb2}, while in strongly quasi-resonant it is given by \eqref{peakb1} and \eqref{peakb3}. 
It is worth noting that all parameters in these formulas should satisfy their respective quasi-resonance conditions ($\rho\approx 0$ or $\rho\approx +\infty$).

\vspace{\baselineskip} 
{{\bf Conflict statement}
	{The authors declare that they have no conflict of interests.}}

\vspace{\baselineskip} 
{{\bf Acknowledgments}
This work is supported by the National Natural Science Foundation of China (Grant No. 12471239), and Guangdong Basic
and Applied Basic Research Foundation (Grant No. 2024A1 515013106)}


\end{document}